\documentclass[preprint,12pt,a4paper]{article}
\usepackage{setspace}
\usepackage{appendix}
\usepackage{graphicx}  
\usepackage{dcolumn}   
\usepackage{bm}        
\usepackage{amssymb}   
\usepackage{amsmath}
\usepackage{array}
\usepackage{epstopdf}
\usepackage{mathtools}
\usepackage[T1]{fontenc}  
\usepackage{textcomp}     
\usepackage{mathtools,slashed}
\usepackage{graphics}
\usepackage{epsfig}
\usepackage{psfrag}
\usepackage{color}
\usepackage[utf8]{inputenc}
\usepackage{scrextend}
\DeclareMathAlphabet{\pazocal}{OMS}{zplm}{m}{n}
\usepackage{amsfonts}
\usepackage{amsthm}
\usepackage{float}
\usepackage{textcomp}
\usepackage[section]{placeins}
\usepackage[caption=false]{subfig}
\usepackage{subfig}
\usepackage{tabularx}

\newcommand*{\rom}[1]

\usepackage{subfig}
\title{Realistic Inflation in Flipped $SU(5)$}
\begin{document}
 	\begin{center}
		\baselineskip 20pt 
		{\Large\bf Realistic Inflation in No-Scale $U(1)_R$ Symmetric Flipped $SU(5)$}
		
		\vspace{1cm}
		
		{\large 
			Mian Muhammad Azeem Abid$^{a,}$\footnote{ azeem$\_92$@live.com}, Maria Mehmood$^{a,}$\footnote{ mehmood.maria786@gmail.com}, Mansoor Ur Rehman$^{a,}$\footnote{mansoor@qau.edu.pk} and Qaisar Shafi$^{b,}$\footnote{qshafi@udel.edu} 
		} 
		\vspace{.5cm}

		{\baselineskip 20pt \it
			$^a$Department of Physics, Quaid-i-Azam University, \\ 
			Islamabad 45320, Pakistan \\
			\vspace{2mm} 
			$^b$Bartol Research Institute, Department of Physics and Astronomy, \\
			University of Delaware, Newark, DE 19716, USA \\
		}
\vspace{1cm}
\end{center}
	
\begin{abstract}
	We have realized non-minimal Higgs inflation and standard hybrid inflation in the supersymmetric flipped $SU(5)$ model with $U(1)_R$ symmetry using the no-scale form of the K\"{a}hler potential. In non-minimal Higgs inflation the waterfall Higgs field plays the role of inflaton, and in standard hybrid inflation the gauge singlet field $S$ is employed as an inflaton. The predictions of both models are in good agreement with the Planck 2018 data. For numerical calculations we have fixed the gauge symmetry breaking scale, $M$, around $2\times 10^{16}$ GeV. In both models the inflaton field values are constrained below $m_P$. The tensor to scalar ratio $r$ in non-minimal inflation is of the order of $10^{-3}$ and for standard hybrid inflation $r$ is tiny, of order $10^{-15} - 10^{-4}$. The scalar spectral index in both cases lie within the Planck 1-$\sigma$ bounds, and the running of the scalar spectral index lies in the range, $-dn_s/d\ln k \sim 6\times 10^{-4}$ for non-minimal model and $10^{-9} - 10^{-3}$ for the standard hybrid model. A realistic scenario of reheating and non-thermal leptogenesis is employed with reheat temperature $T_r \sim 10^9$ GeV for non-minimal model and $10^{6} - 10^{10}$~GeV for standard hybrid model. The $R$-symmetry plays a vital role in forbidding rapid proton decay, but at the same time it also suppresses terms responsible for generating right handed neutrino masses. A realistic scenario of right handed neutrino masses is obtained by considering effective $R$ symmetry breaking at the nonrenormalizable level with adequate suppression of rapid proton decay.
\end{abstract}

\section{Introduction}
Supersymmetric grand unification has a number of well-known attractive features that makes it a compelling extension of the Standard Model (SM). Although LHC Run-2 has not found any evidence for low scale supersymmetry, we remain optimistic that some evidence will appear in the near future, perhaps at the LHC Run-3. Supersymmetry (susy) also provides an attractive framework for implementing the inflationary scenario. A prominent example of this is minimal supersymmetric hybrid inflation in which inflation is associated with the breaking of some local gauge symmetry $G$ \cite{Dvali:1994ms,Copeland:1994vg,Linde:1997sj,Senoguz:2004vu,Rehman:2009nq,Buchmuller:2014epa}. In the simplest example based on $G = U(1)_{B-L}$, say, the temperature  fluctuation ($\delta T / T$) is roughly proportional to $(M / m_P)^2$ \cite{Dvali:1994ms}, where $M$ and $m_P$ respectively denote the gauge symmetry breaking and reduced Planck scales. The inclusion in the inflationary potential of both radiative corrections and soft supersymmetry breaking terms \cite{Senoguz:2004vu,Rehman:2009nq} ensure that the predictions of this model are fully compatible with the Planck 2018 results \cite{Planck:2018vyg,Planck:2018jri}.
Since $M$ in the simplest inflationary models with minimal superpotential and K\"ahler potential turns out to be comparable in magnitude to the supersymmetric grand unification scale $M_{G}\equiv 2\times 10^{16}$~GeV, they have been extended to include realistic supersymmetric groups including $SU(5)$ \cite{Georgi:1972cj} and flipped $SU(5)$ \cite{Barr:1981qv,Derendinger:1983aj}.

An inflation model with a scalar field  non-minimally coupled to gravity is equivalent to the  Starobinsky $\mathcal{R}^2$ model \cite{Starobinsky:1980te}, which is in good agreement with the Planck 2018 results \cite{Planck:2018vyg,Planck:2018jri}. Non-minimal Higgs inflation in non-supersymmetric models is discussed in \cite{Bezrukov:2007ep,Bezrukov:2008ej,DeSimone:2008ei,Barvinsky:2009fy,Okada:2009wz,Okada:2010jf,Linde:2011nh,Okada:2011en,Pallis:2014cda,Bostan:2018evz,Bostan:2019fvk}, and non-minimal SM-like Higgs inflation in the next to minimal supersymmetric standard model (NMSSM) is discussed in \cite{Einhorn:2009bh,Ferrara:2010yw,Lee:2010hj}. This idea is generalized to canonical superconformal supergravity (SUGRA) models in \cite{Ferrara:2010in}.
For studies employing supersymmetric grand unified theory (GUT) Higgs fields to realize this idea, see \cite{Arai:2011nq,Pallis:2011gr}.

{ Supersymmetric GUT Higgs inflation can be realized in the standard hybrid framework by employing a non-minimal K\"{a}hler potential while inflating either below the vacuum expectation value (vev) or above the vev. However, a successful realization of below the vev inflation requires an extra $Z_n$ symmetry with a nonrenormalizable superpotential \cite{Senoguz:2004ky,Masoud:2019gxx,Rehman:2018gnr}. On the other hand, above the vev inflation can be realized with a renormalizable superpotential but with a special form of the K\"{a}hler potential usually employed in no-scale supergravity. A realization of the Starobinsky model in no-scale supergravity \cite{Cremmer:1983bf,Ellis:1983sf,Lahanas:1986uc}
is discussed in \cite{Ellis:2013xoa}. For some recent papers on GUT based inflation in no-scale supergravity framework,S see \cite{Ellis:2017jcp,Ahmed:2018jlv,Ahmed:2021dvo}}.

{Our goal for this paper is twofold. We implement both supersymmetric hybrid inflation as well as non-minimal Higgs inflaton in the flipped SU(5) model based on a no-scale supergravity framework.  We show that in both cases the basic inflationary predictions are compatible with the Planck 2018 results \cite{Planck:2018vyg,Planck:2018jri}. The models are realistic since we discuss reheating followed by an explanation of how the observed baryon asymmetry in the universe arises via leptogenesis. The inflaton field values in our analysis are constrained to lie below the reduced Planck mass, $m_P$, which makes it possible to control the supergravity corrections.  	
An attractive feature of $R$ symmetric flipped $SU(5)$ model is that it naturally solves the  doublet-triplet problem via the missing partner mechanism \cite{Antoniadis:1987dx} without assuming fine tunning. Since the electroweak Higgs doublets remain massless, the Giudice-Masiero mechanism \cite{Giudice:1988yz} is employed for generating the  MSSM $\mu$-term. The latter plays an important role in $\mu$-hybrid inflation, see \cite{Okada:2015vka,Rehman:2017gkm,Wu:2016fzp,Okada:2017rbf,Lazarides:2020zof}. The $R$ symmetry also forbids rapid dimension-four and also dimension-five proton decay operators \cite{Mehmood:2020irm}.}

{The paper is organized as follows. In section~\ref{fsu5} we present the necessary basic ingredients of the $U(1)_R$ symmetric flipped $SU(5)$ model. Section~\ref{model1} is devoted to non-minimal Higgs inflation and its predictions. Section~\ref{rt} deals with the reheat temperature, non-thermal leptogenesis and gravitino constraint. Standard hybrid inflation in the no-scale supergravity framework is studied in section~\ref{model2}. Rapid proton decay in both models is discussed in section~\ref{pdecay}, and our conclusions are summarised in section~\ref{conclusion}. }

\section{$U(1)_R$ Symmetric Flipped $SU(5)$ Model\label{fsu5}}
In the flipped $SU(5)$ model \cite{Barr:1981qv,Derendinger:1983aj,Antoniadis:1987dx} based on the gauge symmetry $SU(5) \times U(1)_X$, the matter superfields ($Q_i,\,U_i^c,\,D_i^c,\,L_i,\,E_i^c$) of MSSM along with a right handed neutrino superfield ($N_i^c$) reside in $10_i (Q_i+D_i^c+N_i^c)$, $\overline{5}_i(U_i^c+L_i)$ and $\overline{1}_i(E_i^c)$ representations, while the Higgs sector consists of a 10-plet pair $10_H (Q_H +D_H^c+N_H^c)\oplus\overline{10}_H (\overline{Q}_H +\overline{D}_H^c + \overline{N}_H^c)$ and a 5-plet pair $5_h (D_h+H_d)\oplus\overline{5}_h (\overline{D}_h+H_u)$. Here, the superfields ($D_h,\,\overline{D}_h$) represent the color triplet scalars, and ($H_u,\,H_d$) are the MSSM Higgs doublets.  The $X$ charge assignments of the various superfields are as follows:
	  \begin{equation}
	  X(10_H, \overline{10}_H, 5_h, \overline{5}_h, 10_i, \overline{5}_i, 1_i)= (1,-1,-2,2,1,-3,5).
	  \end{equation} 
With an additional gauge singlet superfield $S$, we can write the $U(1)_R$ symmetric superpotential of flipped $SU(5)$ model  as \cite{Kyae:2005nv,Rehman:2009yj}, 
\begin{align}
W &= \kappa S ( 10_H\overline{10}_H - M^2)\notag\\
&+\lambda_1 10_H 10_H 5_h + \lambda_2 \overline{10}_H\overline{10}_H\overline{5}_h \notag\\
 &+ y_{i j}^{(d)} 10_i 10_j 5_h + y_{i j}^{(u,\nu)} 10_i \overline{5}_j \overline{5}_h + y_{i j}^{(e)} 1_i \overline{5}_j 5_h, \label{sup}
 \end{align}
where the $R$ charge assignments of the superfields are given in Table \ref{rcharge}.

Here, $q$ is the $R$ charge of $10_H$ superfield which is arbitrary and can be taken as $q=1$. The first term in $W$ with coupling $\kappa$ is relevant for inflation with $S$ and the gauge singlet components ($N_H,\,\overline{N}_H^c$) of Higgs 10-plets playing the role of inflaton respectively in standard hybrid and non-minimal Higgs inflation. The form of this term requires that $R(S)=R(W)=1$ and $R(10_H)=-R(\overline{10}_H)$. Thus, it forbids the dangerous $S^2$ and $S^3$ terms which could generate the well-known eta problem commonly encountered in supergravity inflation models. However, as discussed later, these terms can assist in the stabilization of $S$ in non-minimal Higgs inflation with natural values of the relevant parameters. 
\begin{table}[ht!]
\begin{center}
\caption{\label{rcharge} $R$ charge assignments of superfields}
\begin{tabular}{|c|c|c|c|c|c|c|c|}
\hline
$S$ & $10_H$ & $\overline{10}_H$ & $5_h$ & $\overline{5}_h$ & $10_i$ & $ \overline{5}_i$& $1_i$ \\
\hline
$1$& $q$& $-q$& $1-2q$& $1+2q$& $q$& $-3q$& $5q$\\
\hline
\end{tabular}
\end{center}
\end{table}
\FloatBarrier
 
 The second and third terms with couplings $\lambda_1$ and $\lambda_2$ solve the doublet-triplet splitting problem by mixing the color triplet Higgs pairs $(D_h, \overline{D}_h)$ and $(D_H^c, \overline{D}_H^c)$ and thereby providing them with GUT scale masses. These terms require that $R(5_h)= 1-2 R(10_H)$, $R(\overline{5}_h) = 1+2 R(10_H))$ and $R(5_h\overline{5}_h) = 2 \neq R(W)$. Thus, the $R$ symmetry forbids the appearance of the term $5_h\overline{5}_h$ and avoids potential GUT scale masses for the MSSM doublets. However, to solve the MSSM $\mu$ problem, the Giudice-Masiero mechanism \cite{Giudice:1988yz} is assumed. The remaining terms in the superpotential (Eq.~(\ref{sup})) with couplings $ y_{i j}^{(d)}$, $ y_{i j}^{(u,\nu)}$ and $ y_{i j}^{(e)}$ are Yukawa terms responsible for generating the masses of quarks and leptons. The $R$ charge assignments for these terms are $ R(10_i) = R(10_H)$, $ R(\overline{5}_i) = -3 R(10_H)$ and $R(1_i) = 5 R(10_H)$.

  The relation $ R(10_i) = R(10_H)$ forbids the term, $\overline{10}_H \overline{10}_H 10_i 10_j /m_P$ which is important for generating the light neutrino masses via the seesaw mechanism. Although the light neutrino masses could be generated with non-minimal field  contents \cite{Kyae:2005nv}, we prefer to employ the minimal framework mentioned above and assume $R$ symmetry breaking at nonrenormalizable level in the superpotential (Eq.~(\ref{sup})) \cite{Civiletti:2013cra}. As discussed in a later section, this term can also play an important role in proton decay, reheating and baryogenesis via leptogenesis. Finally, the $Z_2$ subgroup of $U(1)_R$ cannot play the role of $R$ parity as $ R(10_i) = R(10_H)$. Therefore, an additional unbroken $Z_2$ symmetry is assumed to ensure the stability of the lightest supersymmetric particle (LSP) which can serve as a potential candidate for cold dark matter. Under this $Z_2$ parity, all matter superfields ($10_i,\,\overline{5}_i,\,1_i$) are odd whereas all Higgs superfields including $S$ are even.  This symmetry also avoids many unwanted terms including $10_H \overline{5}_h \overline{5}_i \supset \langle N_H^c\rangle H_u L_i$ and $10_H 5_h 10_i \supset \langle N_H^c\rangle \overline{D}_H^c D_i^c$ which can assign superheavy masses to $H_u$, $L_i$ and $D_i^c$ \cite{Kyae:2005nv}.

\section{Non-Minimal Higgs Inflation}\label{model1}
In non-minimal Higgs inflation the SM gauge singlet component of the $10$-plet Higgs pair (along the $D$-flat direction) plays the role of inflaton. To realize non-minimal Higgs inflation in flipped $SU(5)$ with $R$ symmetry, we assume the following no-scale like form of the K\"{a}hler potential,
     \begin{eqnarray} \label{K}
          K &=& -3 \log  \left( 1 -\frac{1}{3}\left( \lvert S \rvert ^2 + \lvert 10_H \rvert ^2+ \lvert \overline{10}_H \rvert ^2 + \lvert 5_h \rvert ^2+ \lvert  \overline{5}_h \rvert^2 \right. \right. \notag \\
 &&  \left. \left. + \lvert 10_i \rvert ^2 + \lvert \overline{5}_i \rvert ^2 + \lvert \overline{1}_i \rvert ^2 \right)  +  \frac{1}{2} \chi(10_H\overline{10}_H + h.c)+\frac{1}{3}\gamma\lvert S\lvert^4 + \cdots\right),
\end{eqnarray}
where we have assumed the stabilization of the modulus fields \cite{Cicoli:2013rwa, Ellis:2013nxa} and utilized units where the reduced Planck mass, $m_P = 1$. The exact no-scale limit is obtained with vanishing $\chi$ and $\gamma$ couplings. The term with coupling $\chi$ plays a crucial role in non-minimal Higgs inflation, and the higher order term with $\gamma$ coupling is needed to stabilize $S$ during inflation \cite{Lee:2010hj}. However, as discussed later, the $\gamma$-term plays a significant role in realizing standard hybrid inflation by employing the above form of $K$. In order to  obtain MSSM the 10-plet Higgs pair attains a non-zero vev along the $N_H^c$, $\overline{N}_H^c$ directions in the following global susy vacuum, 
     \begin{equation}     
\langle 10_H \overline{10}_H \rangle =\langle N_H^c \overline{N}_H^c \rangle = M^2, \qquad \text{and} \qquad \langle S \rangle = 0,
     \end{equation}
where the gauge symmetry breaking scale $M$ is taken to be  the order of GUT scale, $M_{G}\equiv 2\times 10^{16}$~GeV. It is interesting to note that no other gauge invariant bilinear term appears in the K\"ahler potential mainly due to $R$-symmetry. This highlights the importance of the term, $10_H \overline{10}_H+h.c$, as a unique possibility in generating the required nonminimal coupling in the present model. The realization of no-scale inflation with $R$-symmetry breaking effects are discussed in \cite{Khalil:2018iip,Moursy:2020sit} with $S$ field inflaton and in \cite{Moursy:2021kst}  with sneutrino $N^c$ inflaton.

    The scalar potential in the Einstein' frame is defined as, 
     \begin{equation}
     V_E = e^{G}\left( G_i (G^{-1})_j^i  G^j- 3\right)+\frac{1}{2} g_a^2 G^i (T_a)_i^j z_j (\text{Re} f^{-1}_{ab}) G^k (T_b)_k^l z_l,
     \end{equation}
where
     \begin{equation}
    G = K + \log|W|^2, \quad  G^i \equiv \frac{\partial G}{\partial z_i} ,\quad G_i = \frac{\partial G}{\partial z_i^*}, \quad G^i_j = \frac{\partial G}{\partial z_i \partial z_j^*},
     \end{equation}
with $z_i \in \{S, 5_h, \overline{5}_h, 10_H, \overline{10}_H,\cdots\}$, $f_{ab}$ is the gauge kinetic  function and $T_a$ are the generators of the gauge group. After the stabilization of $S$, $5_h$, $\overline{5}_h$ and the matter fields in their appropriate vacua, the scalar potential in Einstein' frame becomes, 
    \begin{eqnarray}
    V_E &=& \dfrac{\kappa^2 \left| N_H^c \overline{N}_H^c - M^2 \right|^2 }{\left( 1-\frac{1}{3}(|N_H^c|^2+ |\overline{N}_H^c| ^2) + \frac{\chi}{2}(N_H^c \overline{N}_H^c + h.c) \right)^2} \notag \\
    &+& \frac{1}{2} \left(\frac{3}{5}g_5^2 +  g_X^2 \right) \frac{(|N_H^c|^2 - |\overline{N}_H^c| ^2 )^2}{\left( 1-\frac{1}{3}(|N_H^c|^2+ |\overline{N}_H^c| ^2) + \frac{\chi}{2}(N_H^c \overline{N}_H^c + h.c) \right)^2},
    \end{eqnarray}
where we have rotated the 10-plet Higgs fields in the SM gauge singlet neutrino direction using flipped $SU(5)$ gauge invariance and set $f_{ab}=\delta_{ab}$. The gauge couplings $g_5$ and $g_X$ belong to the $SU(5)$ and $U(1)_X$ gauge groups respectively. We can rewrite the complex neutrino fields in terms of real scalar fields as
\begin{equation}
N_H^c = \frac{h}{\sqrt{2}} e^{i\alpha}\cos\beta, \quad \overline{N}_H^c = \frac{h}{\sqrt{2}} e^{i\overline{\alpha}}\sin\beta,
\end{equation}
where along the D-flat direction ($|N_H^c| = |\overline{N}_H^c| $) the phases, $\alpha,\,\overline{\alpha}$ and $\beta$ can be stabilized at
\begin{equation}
\beta = \frac{\pi}{4}, \quad \alpha = \overline{\alpha} = 0.
\end{equation}
     
 \begin{figure}%
\centering
      	{\subfloat[]{\includegraphics[width=2.7in]{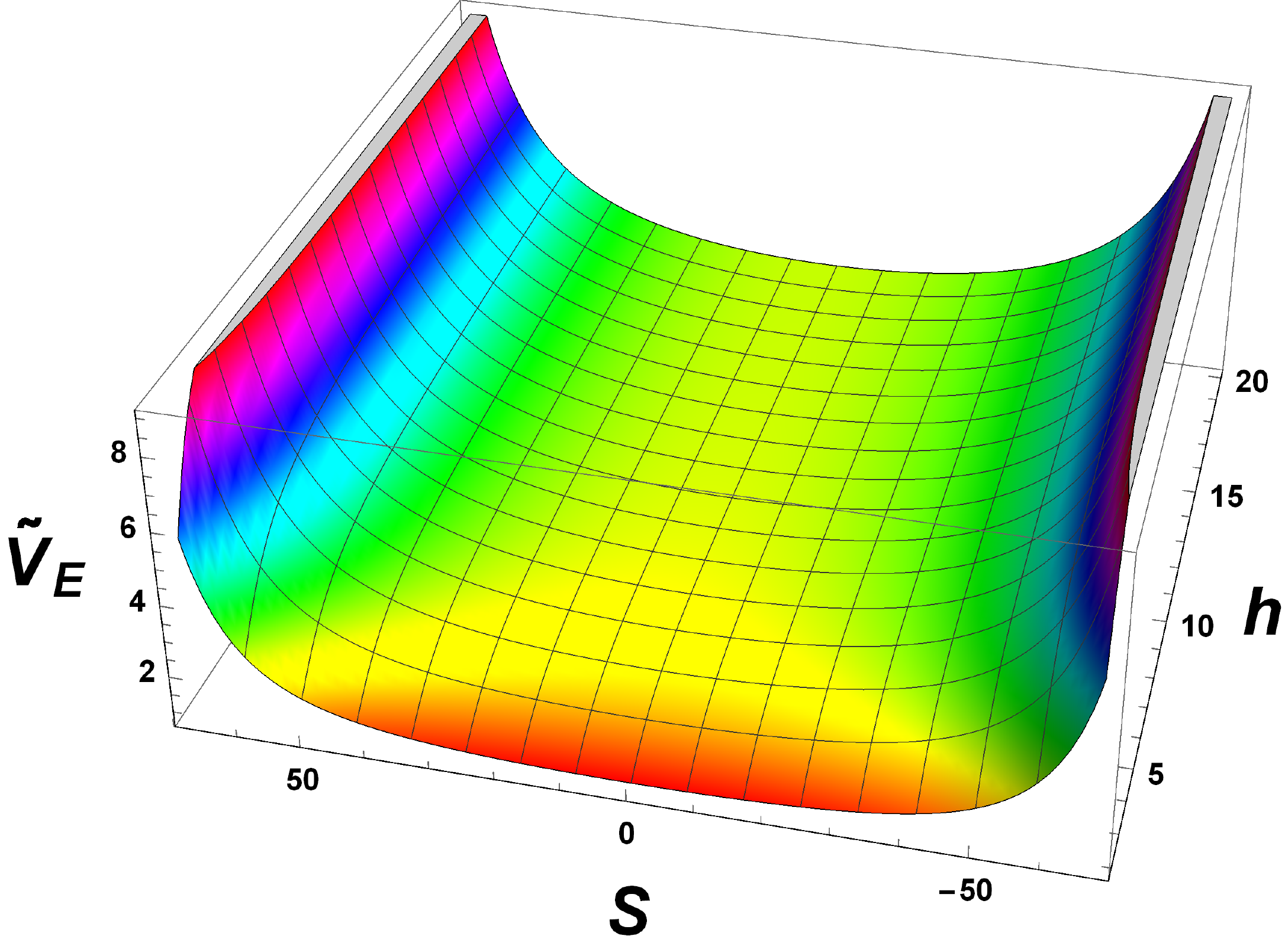} }}\qquad
      	{\subfloat[]{\includegraphics[width=2.7in]{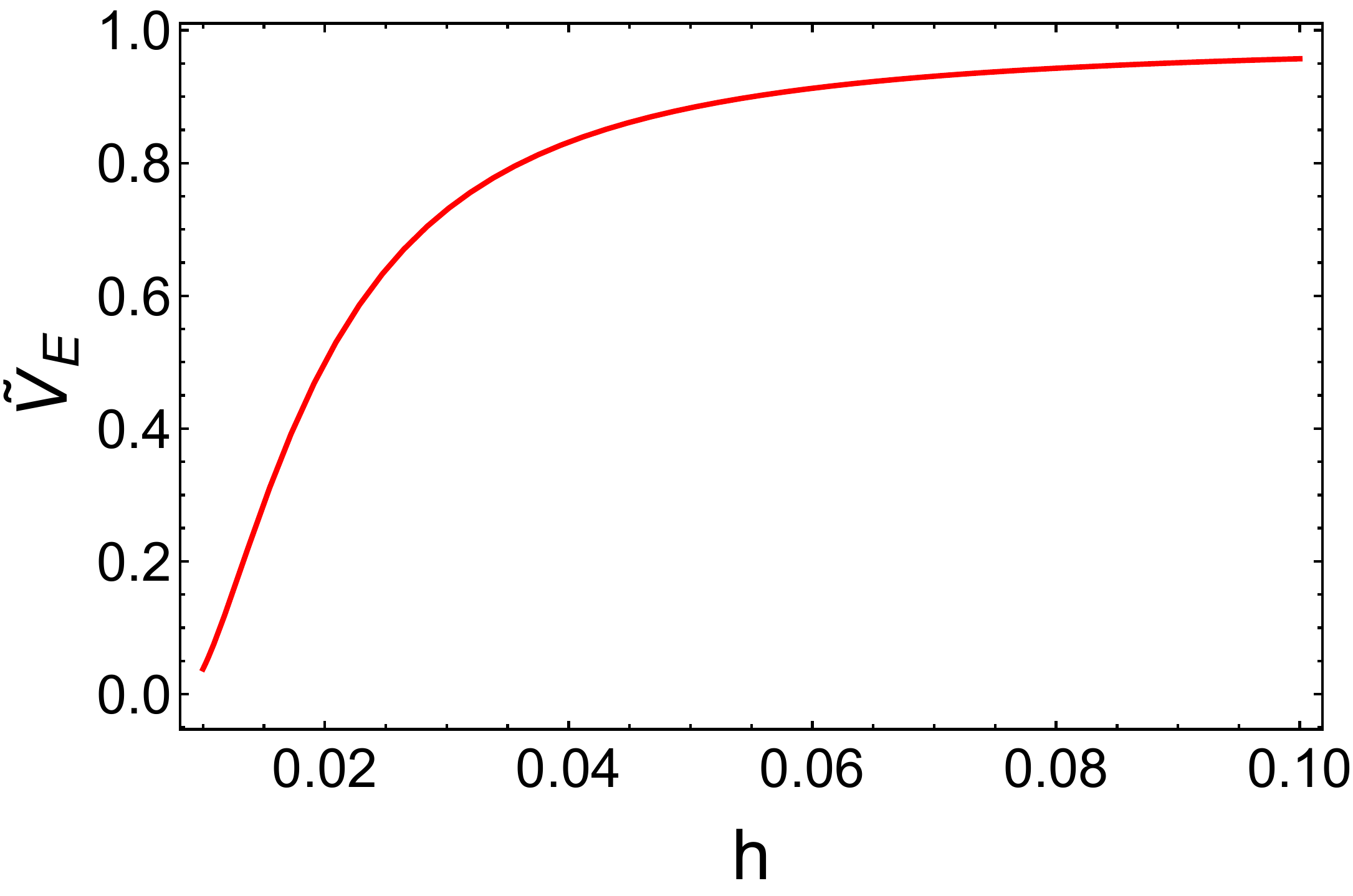} }}
      	\caption{The normalized scalar potential $\tilde{V}_E=V_E(S,h)/V_0$ (left panel) and its $S=0$ slice (right panel)  for $ M =0.0082, \chi= 15000, \gamma=9, \kappa= 0.156$. Here the field values are written in Planckian units and $V_0 = \kappa^2 M^4$.}%
\label{fig:3d}%
\end{figure}     
   
The two-field scalar potential in Einstein' frame along the D-flat direction becomes
\begin{eqnarray}
      V_E(S,h)& \simeq & \frac{\kappa^2 }{16}\frac{(h^2-4M^2)^2}{(1-4\gamma S^2)(1+\xi h^2)^2},\quad \text{ with } 
   \quad \xi \equiv \frac{\chi}{4}-\frac{1}{6}.
\end{eqnarray}
{ {This expression represents a good approximation of the exact scalar potential plotted in Fig.~\ref{fig:3d}}} for typical values of the relevant parameters where the special role of the higher order $\gamma$-term in generating the $S=0$ minimum is obvious. Finally, the single-field scalar potential in Einstein frame takes the following form,
\begin{equation} \label{Epot}
V_E (h)  = \frac{\kappa^2}{16} \frac{\left( h^2 - 4 M^2 \right)^2}{\Omega^2} = \frac{\kappa^2}{16} \frac{\left( h^2 - 4 M^2 \right)^2}{(1+\xi h^2)^2},
   \end{equation}
where the conformal scaling factor, $\Omega$, relating the Einstein' and Jordan' frames is given by
\begin{equation}
g^J_{\mu\nu} = \Omega g^E_{\mu\nu} = (1+\xi h^2) g^E_{\mu\nu}.
\end{equation}
The Lagrangian in the Einstein' frame is
\begin{equation}\label{Jordan}
\mathcal{L}_E=\sqrt{-g_E}\left[\dfrac{1}{2}\mathcal{R}(g_E) - \frac{1}{2} g_E^{\mu \nu} \partial_\mu \hat{h} \partial_\nu \hat{h} - V_E (\hat{h}(h))\right],
   \end{equation}
where $\hat{h}$ is the canonically normalized inflaton field defined as,
\begin{equation}
\frac{d\hat{h}}{dh} \equiv J = \sqrt{\dfrac{\Omega+6\xi^2h^2}{\Omega^2}}.
\end{equation}
Here $\mathcal{R}$ is the Ricci scalar. 
   
The slow-roll parameters can be expressed in terms of $h$ as
\begin{eqnarray}
 \epsilon(h)&=&\frac{1}{2J^2}\left(\dfrac{V_E^\prime (h)}{V_E }\right)^2, \quad
 \eta(h) = \frac{1}{J^2} \left( \frac{V_E^{\prime \prime}(h)}{V_E} -  \sqrt{2 \epsilon} J^{\prime}(h) \right),  \label{epsilon} \\
  \zeta^2(h) &=& \frac{\sqrt{2 \epsilon}}{J^3} \left(\dfrac{V_E^{\prime \prime \prime}(h)}{V_E} - 3 \eta J J^{\prime}(h) - \sqrt{2 \epsilon} J^{\prime \prime}(h)\right), 
\end{eqnarray}
where primes denote the derivatives with respect to $h$. 
The scalar spectral index $n_s$, the tensor to scalar ratio $r$ and the running of the scalar spectral index $\frac{d n_s}{d\ln k}$ are given, to leading order in slow-roll parameters, as
\begin{eqnarray}
 n_s &\simeq& 1-6 \epsilon(h_0) + 2 \eta(h_0), \quad r \simeq 16 \epsilon(h_0), \label{nsr} \\
\frac{d n_s}{d\ln k} &\simeq& 16\epsilon(h_0)\eta(h_0) -24 \epsilon^2(h_0)-2\zeta^2(h_0),  \label{alfa}
\end{eqnarray}
where $h_0$ is the field value at the pivot scale as defined below. The analytic expressions of the various inflationary parameters given below are approximated in the large $|\xi|$ limit with $|\xi| M^2$ finite. For example, with $10^2 \lesssim |\xi|\lesssim 10^4$ we obtain $0.007 \lesssim  |\xi|M^2 \lesssim 1$ for $M=M_{G}$. 

The amplitude of the scalar power spectrum is given by 
\begin{equation}\label{As}
A_s(k_0)=\left. \frac{1}{{24} \pi^2 }\frac{V_E(h)}{\epsilon(h)}\right|_{h(k_0) = h_0},
\end{equation}
which is normalized to be $A_s(k_0) = 2.137\times10^{-9}$ at the pivot scale $k_0=0.05\,\text{Mpc}^{-1}$ by the Planck 2018 data \cite{Planck:2018jri}. This normalization constraint can be used to express $\kappa$ in terms of $h_0$ as
\begin{equation} \label{kxi1}
\kappa \simeq \frac{16 \pi \sqrt{2 A_s(k_0)}  ( 1 + 4 \xi M^2 ) (1 + \xi h_0^2)}{|\xi|(h_0^2 - 4 M^2)^2}.
\end{equation} 
The last $N_0$ number of e-folds from $h=h_0$ to the end of inflation at $h=h_e$ is expressed as 
\begin{equation} \label{N0}
N_0=\int_{h_e}^{h_0} dh \dfrac{J(h)}{\sqrt{2\epsilon(h)}}.
\end{equation}
The field value $h_0$ can now be expressed  in terms of $N_0$ as 
 \begin{equation}  \label{h0}
  h_0 \simeq \sqrt{h_e^2 + \frac{4 N_0 (1 + 4 \xi M^2 )}{3 \xi}},
 \end{equation}
where the field value $h_e$ is obtained from
\begin{equation}
\epsilon(h_e)=1  \quad \Rightarrow  \quad  h_e \simeq \sqrt{4 M^2 + \frac{ 2(1 + 4 \xi M^2)}{\sqrt{3}\,\xi}}.
\end{equation}
For numerical predictions of the various inflationary parameters we use the following expression for $N_0$ \cite{Kolb:1990vq},
\begin{equation} \label{Nth}
N_{0}\simeq 53 + \frac{1}{3} \ln\left(\frac{T_{r}}{10^{9}\text{ GeV}}\right)+\frac{2}{3}\ln\left(  \frac{\sqrt{\kappa} M}{10^{15} \text{ GeV}}\right),
\end{equation}
assuming a standard thermal history of Universe.

\begin{figure}[t!]
	\centering
	\subfloat[]{\includegraphics[width=3.033in]{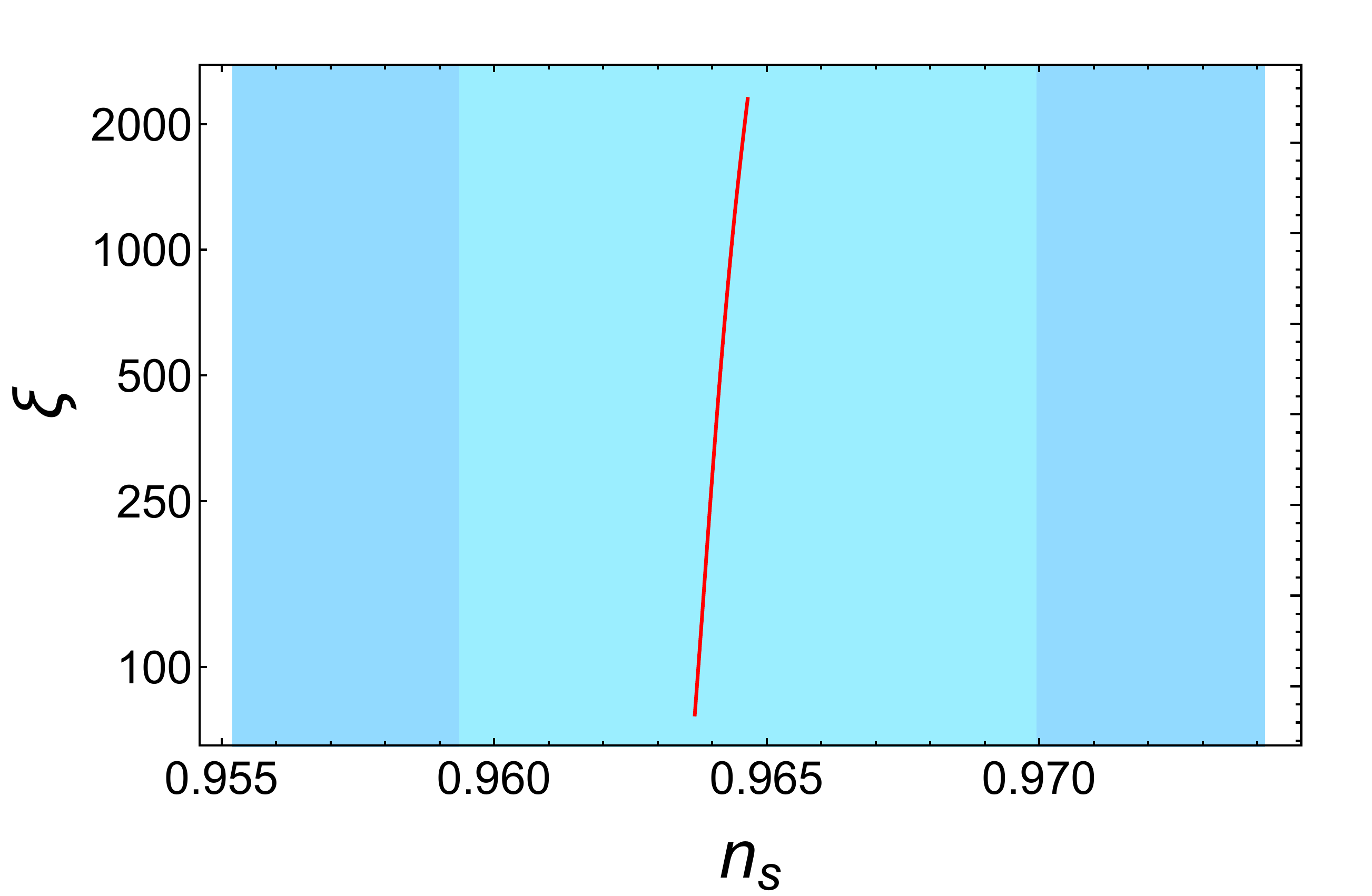}}\ 
	\subfloat[]{\includegraphics[width=2.919in]{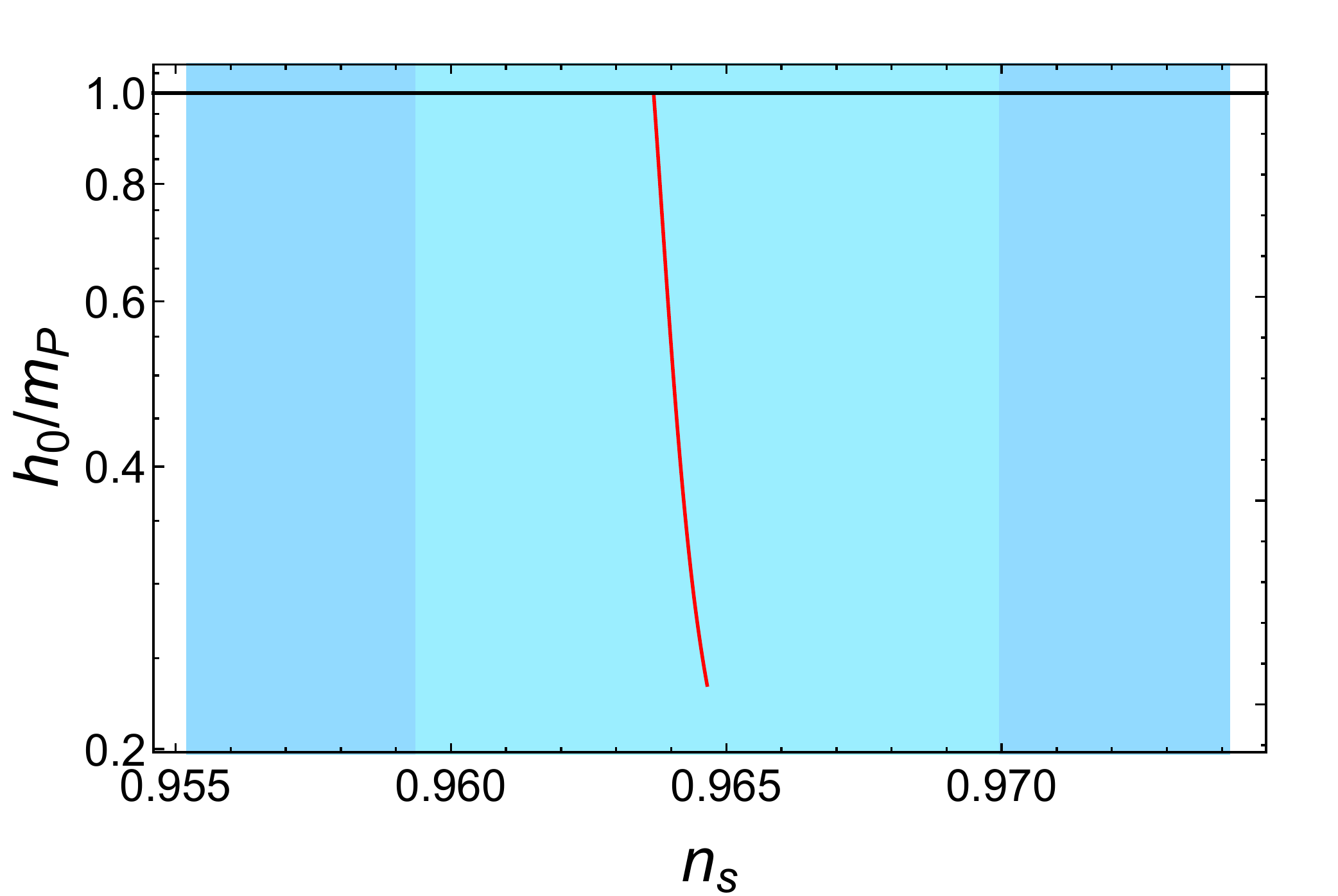}} 
	\caption{The non-minimal coupling $\xi$ and the inflaton field value $h_0$ at the pivot scale versus the scalar spectral index $n_s$. The lighter and darker blue bands represent the Planck 1-$\sigma$ and 2-$\sigma$ bounds respectively.}
\label{fig:xih0}
\end{figure}

The scalar spectral index $n_s$ can be expressed in terms of $h_0$ as
 \begin{equation}\label{ns}
 n_s \simeq 1 - 2 \dfrac{(1 + 4 \xi M^2)\left[ 4 M^2 - 3(1 + 4 \xi M^2)h_0^2 -12(1 + 8 \xi M^2)h_0^4 + 12 \xi^3 h_0^6 \right]}{9\,\xi^4 h_0^4\left(h_0^2 - 4 M^2\right)^2}.
\end{equation}
For given $N_0$ and $M$, the above analytic expressions of $n_s$ (in Eq.~(\ref{ns})) and $h_0$ (in Eq.~(\ref{h0})) as a function of $\xi$ represent a valid approximation of the numerical results displayed in Fig.~\ref{fig:xih0}. In the large $\xi$ limit with $|\xi|M^2 \ll 1$, we obtain the well-known results \cite{Bezrukov:2007ep,Okada:2010jf},
\begin{equation}
(h_0,\,h_e) \rightarrow \left(\sqrt{\frac{4N_0}{3 \,\xi}},\, \left( \frac{4}{3} \right)^{1/4} \frac{1}{\sqrt{\xi}} \right)
\quad \text{and} \quad n_s \rightarrow 1 - \frac{2}{N_0}.
\end{equation}
Requiring the inflaton field value to be sub-Planckian (i.e., $h_0<m_P$), we obtain a lower bound on $\xi \lesssim 4 N_0/ 3$. 

\begin{figure}[t!]
\centering
\subfloat[\label{fig:ralfa1}]{\includegraphics[width=3.05in]{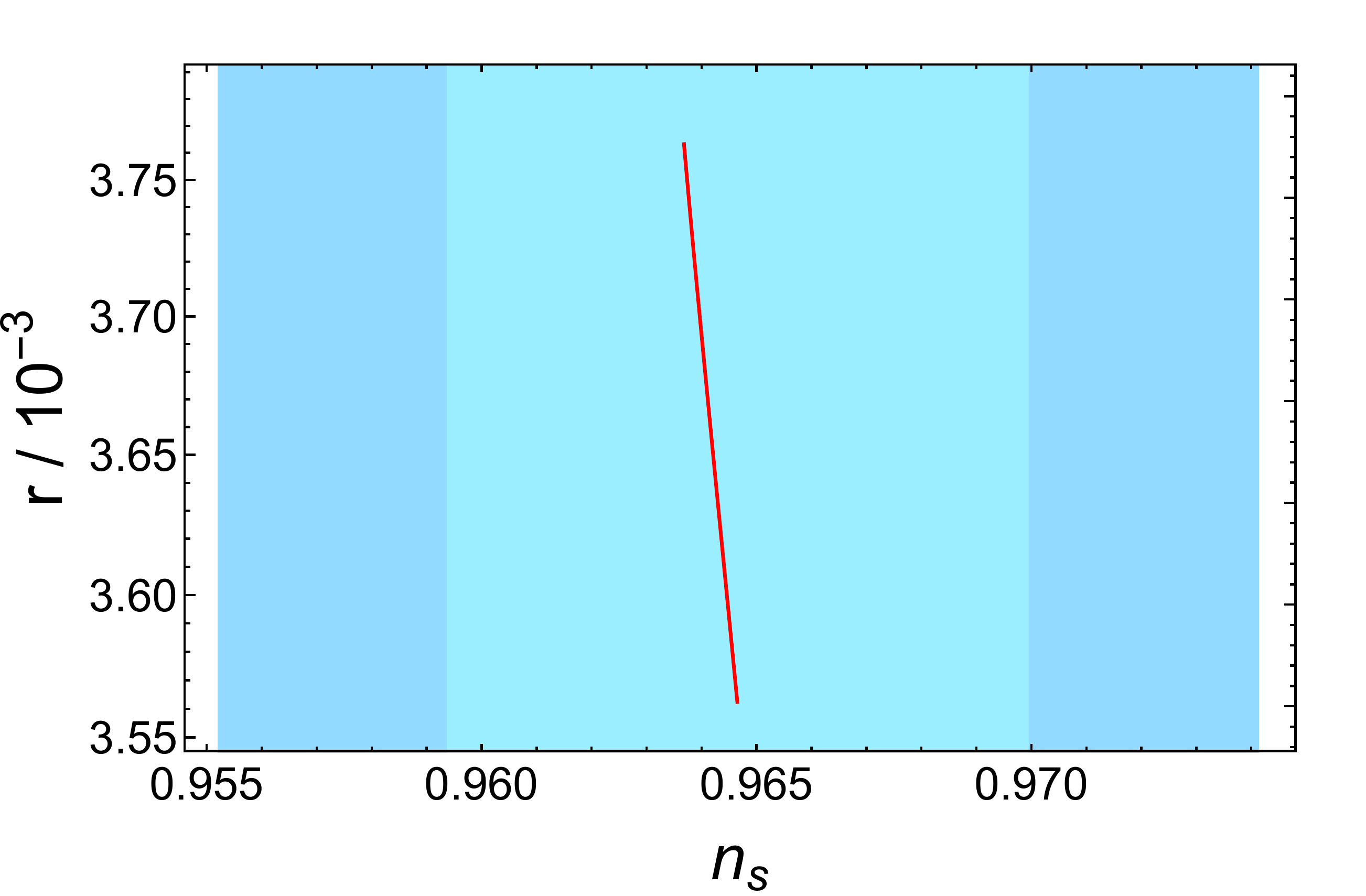}}
\subfloat[\label{fig:ralfa2}]{\includegraphics[width=3.01in]{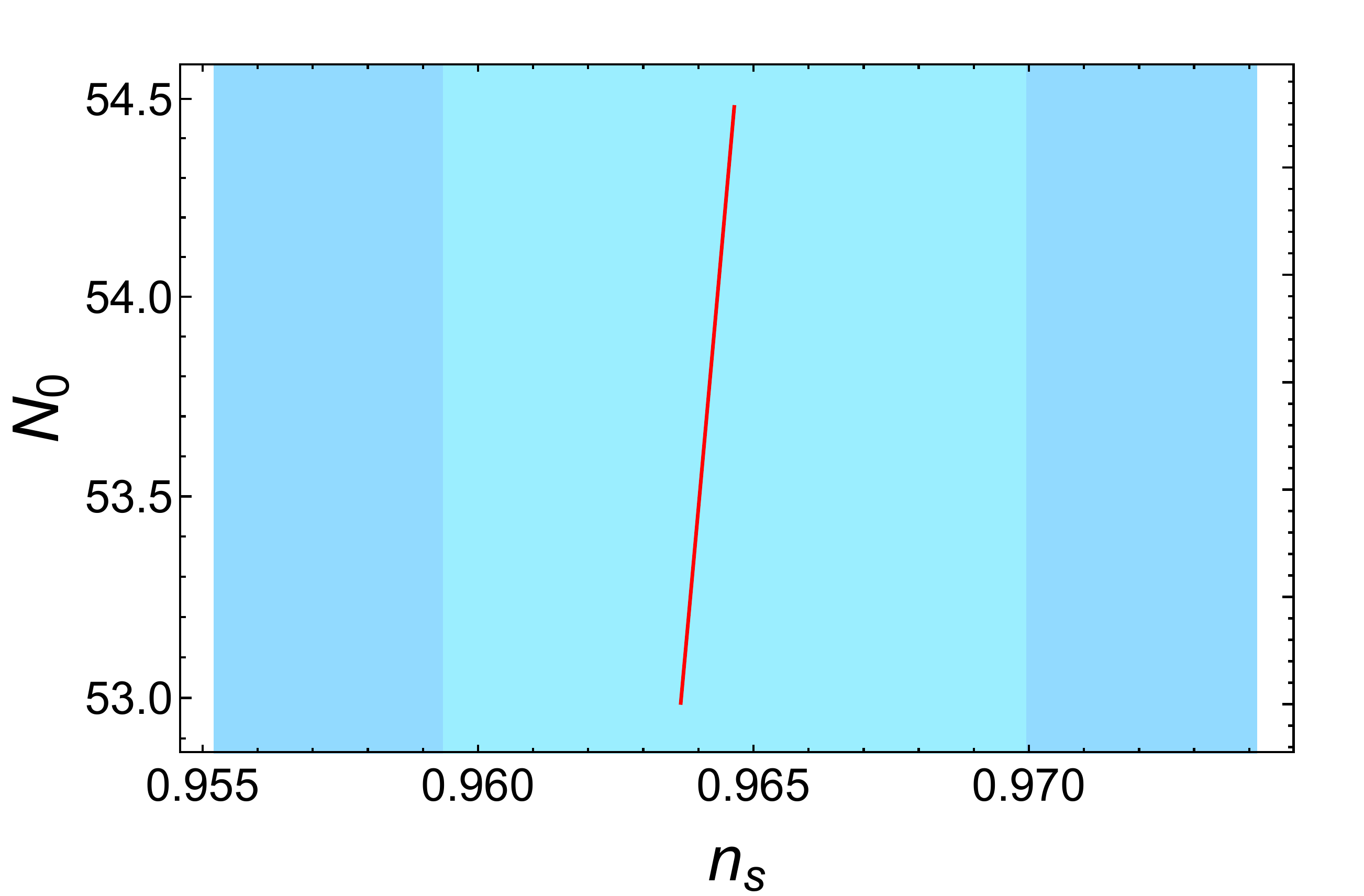}}
\caption{The tensor to scalar ratio $r$ and the number of e-folds $N_0$ versus the scalar spectral index $n_s$. The lighter and darker blue bands represent the Planck 1-$\sigma$ and 2-$\sigma$ bounds respectively.}
\label{fig:ralfa}
\end{figure}

\begin{figure}[t!]
\centering
\subfloat[\label{fig:kxinsa}]{\includegraphics[width=3in]{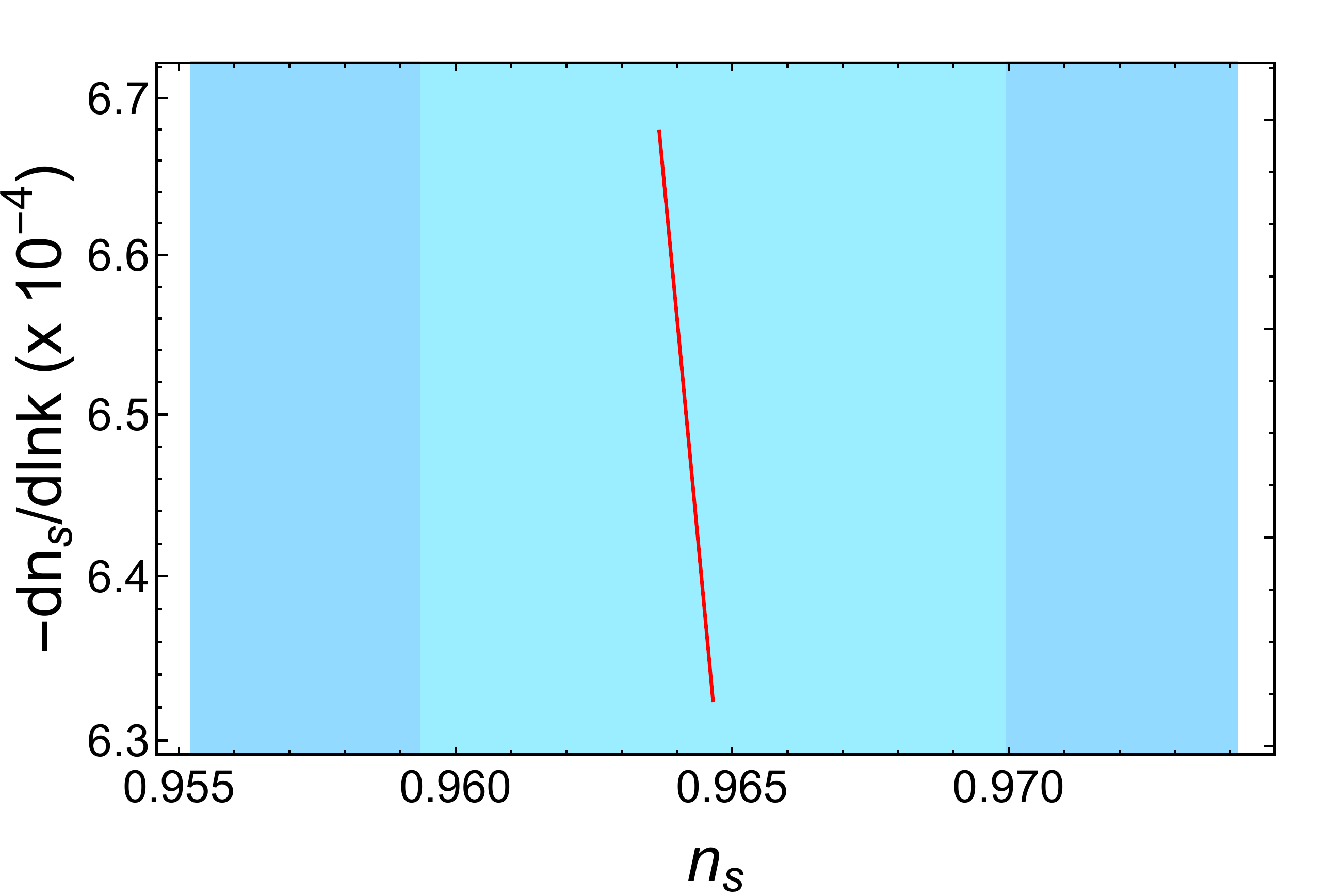}} 
\subfloat[\label{fig:kxinsb}]{\includegraphics[width=3in]{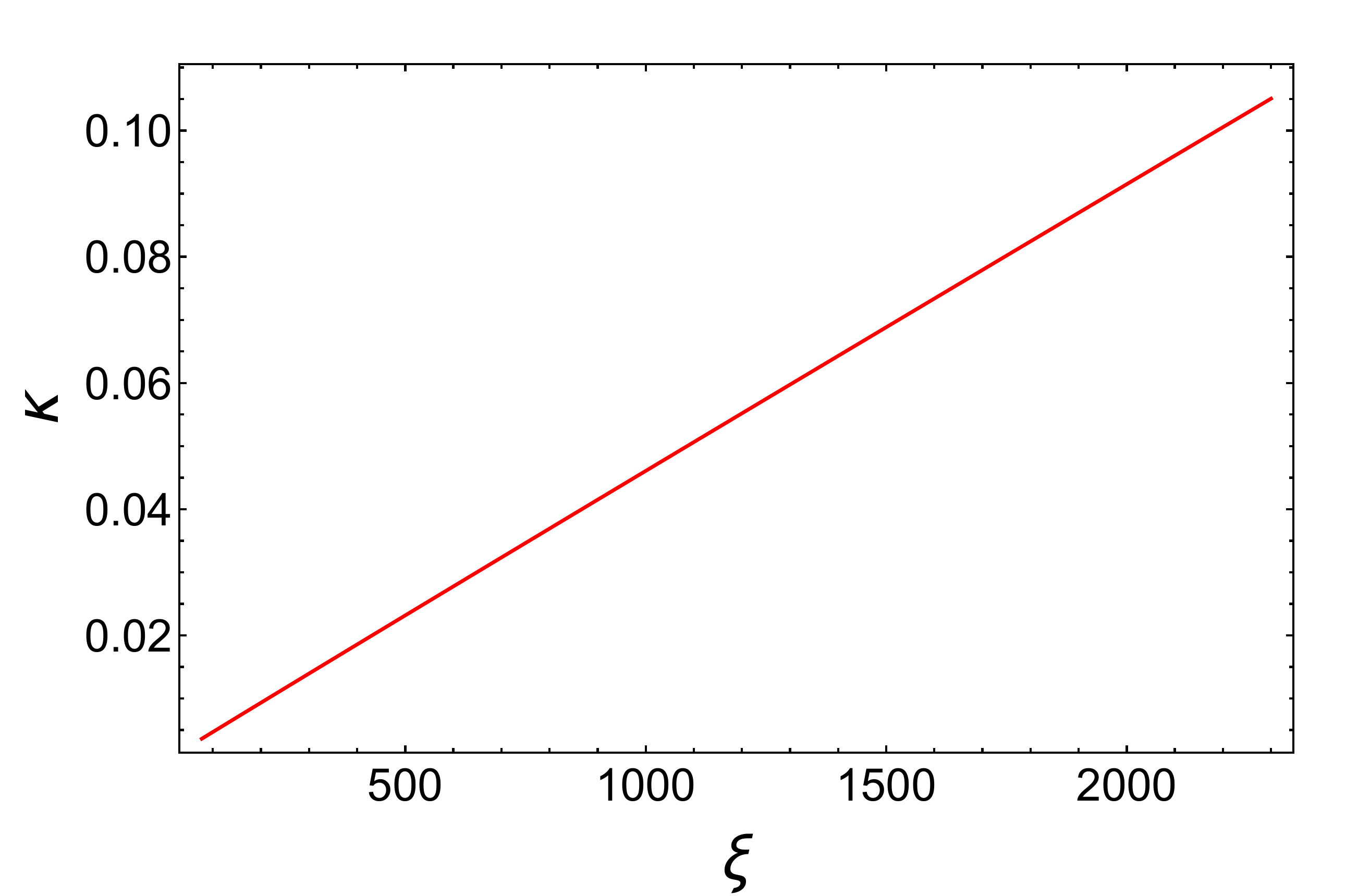}} 
\caption{Running of the scalar spectral index $d\ln{n_s}/dk$ versus the scalar spectral index $n_s$ (left panel), and $\kappa$ versus $\xi$ (right panel). The lighter and darker blue bands represent the Planck 1-$\sigma$ and 2-$\sigma$ bounds respectively.}
\label{fig:kxins}
\end{figure}

In the large $\xi$ limit with $|\xi|M^2 \ll 1$, the tensor to scalar ratio, $r$, reduces to \cite{Bezrukov:2007ep,Okada:2010jf}
\begin{equation}\label{r}
r\simeq \frac{12}{N_0^2}.
\end{equation}
As displayed in the Fig.~\ref{fig:ralfa1}, the value of $r$ varies between $0.0035$ and $0.0037$ and is consistent with the range $N_0 \simeq 52.9-54.4$ shown in the Fig.~\ref{fig:ralfa2}. This range of $N_0$ is obtained, via Eq.~(\ref{Nth}), with the reheating scenario discussed in next section. The range of $r$ obtained here is typical for non-minimal Higgs inflation in large $\xi$ limit.

The running of the scalar spectral index $dn_s/d \ln k$ is approximately given as
\begin{equation} \label{zeta2}
\dfrac{dn_s}{d \ln k} \simeq  -\frac{r(1-n_s)}{2}\simeq -\frac{12}{N_0^3} \simeq -10^{-3},
\end{equation}
in the large $\xi$ limit as shown in the Fig.~\ref{fig:kxinsa}. Finally, the variation of $\kappa$ with respect to $\xi$ is shown in the Fig.~\ref{fig:kxinsb}. In the large $\xi$ limit the following analytic expression of $\kappa$ in terms of $\xi$ can be deduced from Eq.~(\ref{kxi1}),
\begin{equation} \label{kxi}
\kappa \simeq\left(\dfrac{3\pi \sqrt{2 A_s}}{N_0}\right)\xi,
\end{equation}
i.e., $\kappa$ is directly proportional to $\xi$. Therefore, for $77 \lesssim |\xi| \lesssim 2300$ we obtain $0.0036 \lesssim \kappa \lesssim 0.1$, which matches with the numerical estimates shown in the Fig.~\ref{fig:kxinsb}. For this range of $\kappa$, the radiative corrections in standard hybrid inflation play an active role. However, as we have checked explicitly, the radiative corrections along with other soft susy breaking terms are negligible in non-minimal Higgs inflation.

\section{Reheat Temperature, Non-thermal Leptogenesis and Gravitino Constraint\label{rt}}
\begin{figure}[t!]
\centering
\subfloat[\label{minfk}]
{\includegraphics[width=3.14in]{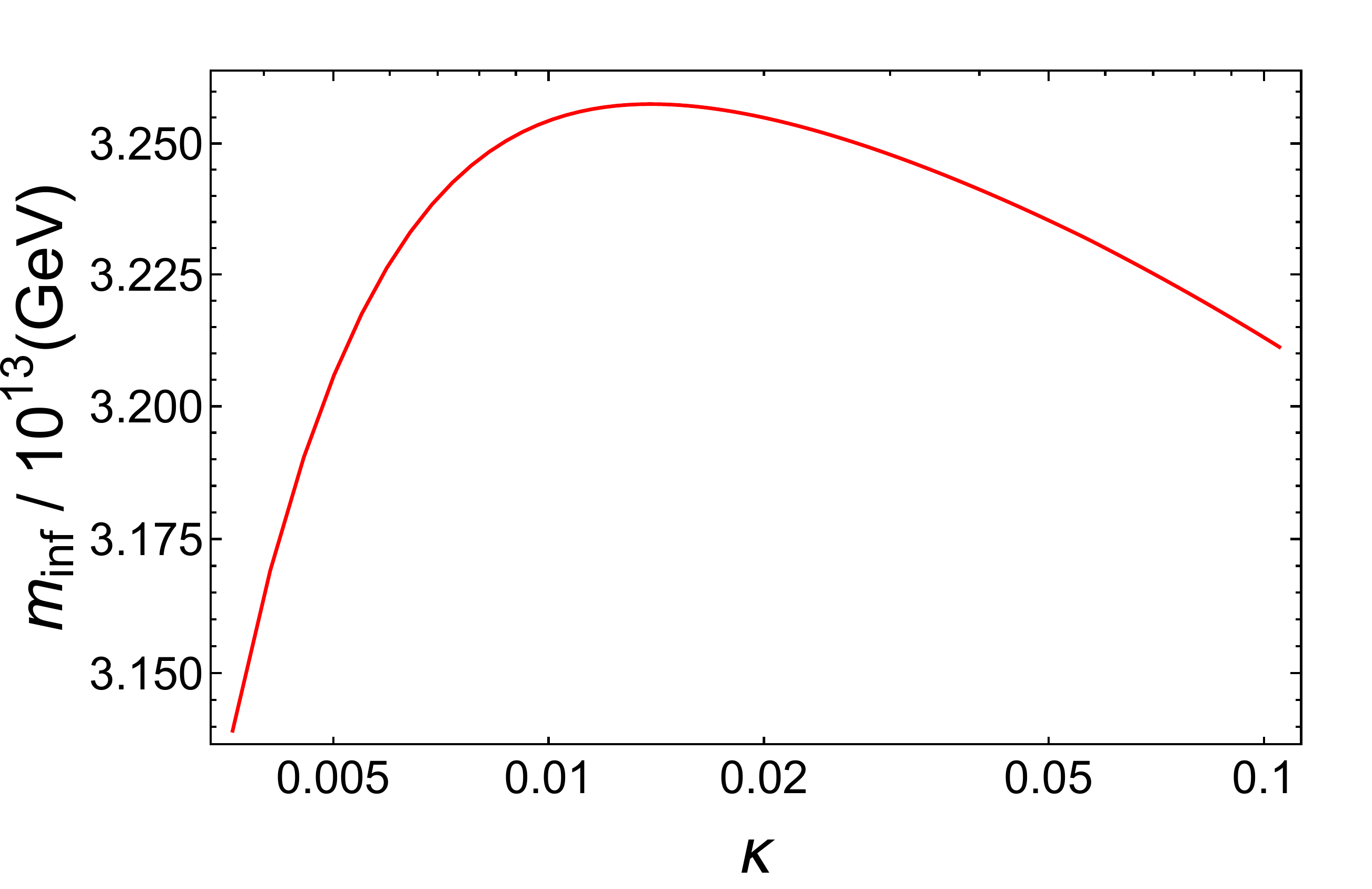}}
\subfloat[\label{mvk}]{\includegraphics[width=3.09in]{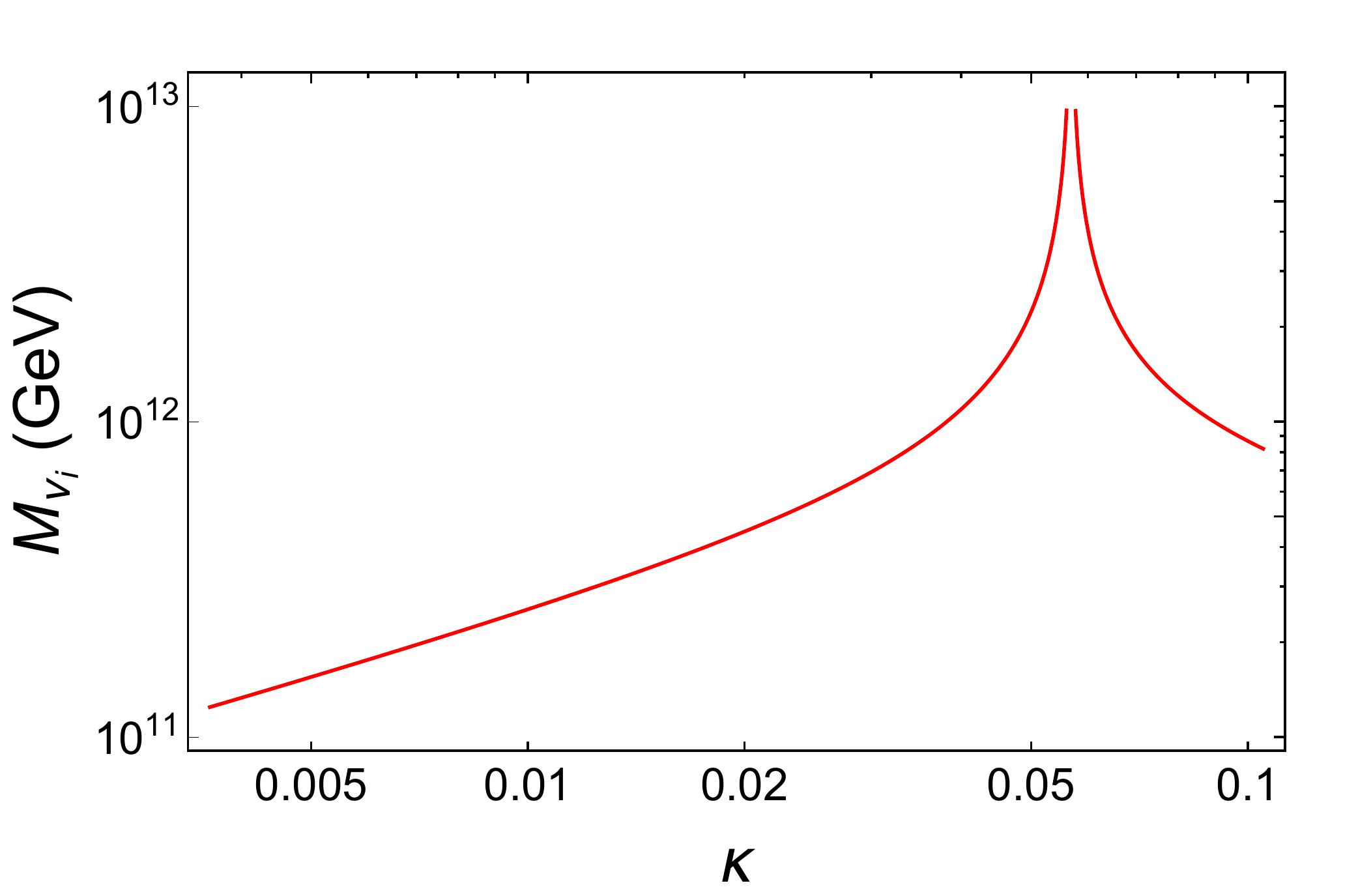}}
\\
\subfloat[\label{trk}]
{\includegraphics[width=2.97in]{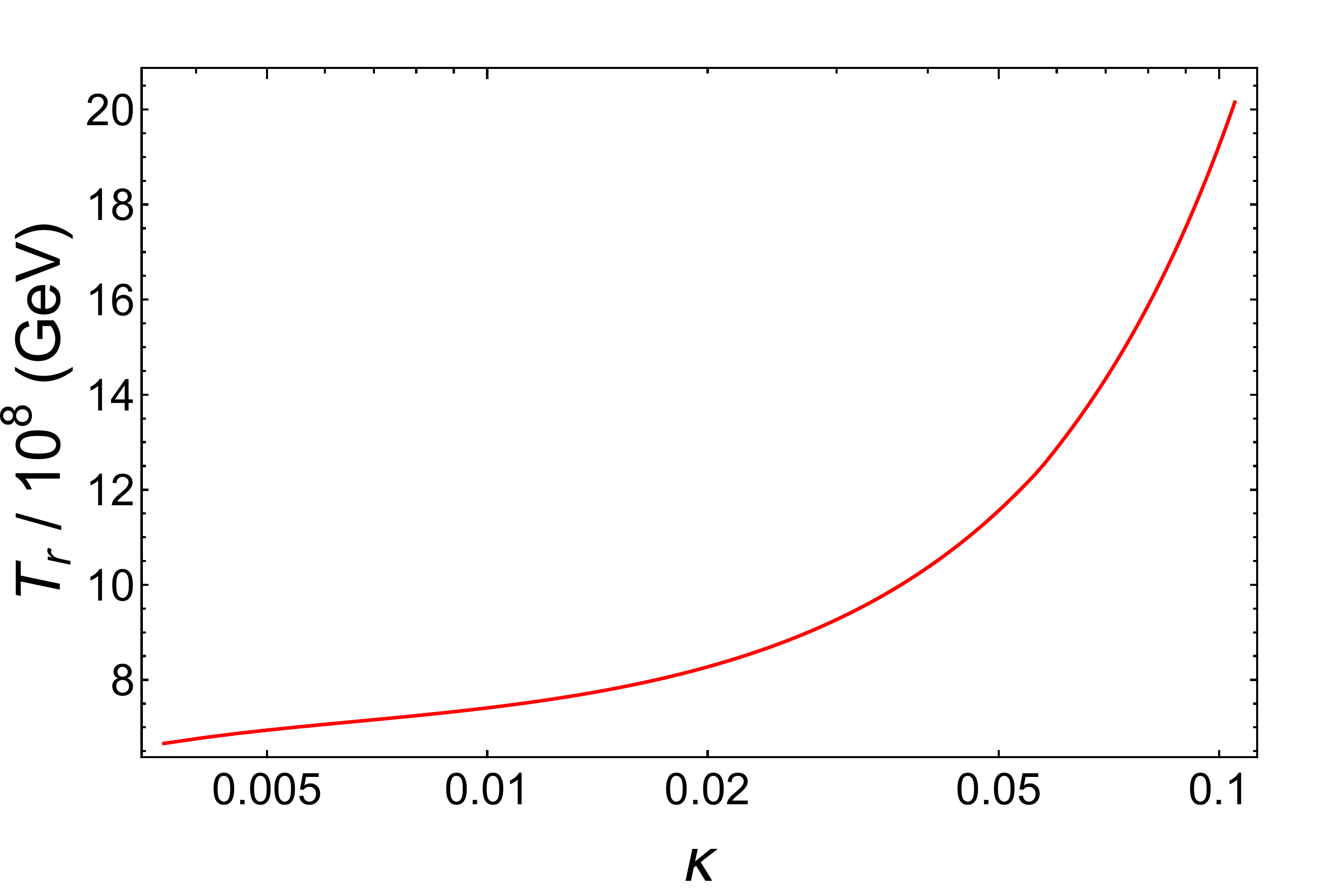}} 
\subfloat[\label{mvtrk}]{\includegraphics[width=3.1in]{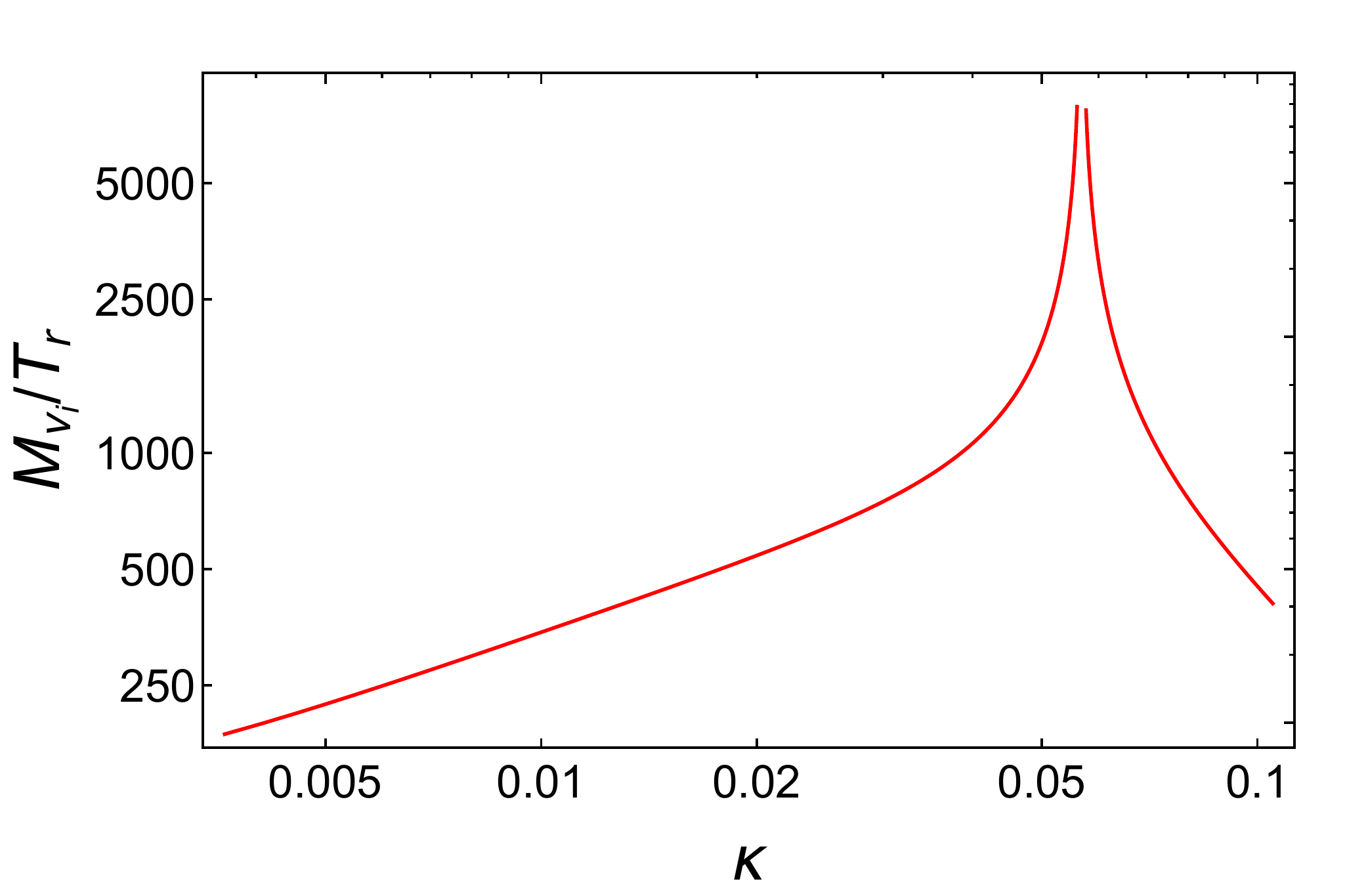}}
\caption{The inflaton mass $m_{\text{inf}}$, the lightest right handed neutrino mass $M_{\nu_i}$, the reheat temperature $T_r$ and the ratio $M_{\nu_i}/T_r$ versus $\kappa$.}
\label{fig:minfmnuvsk}
\end{figure}

With inflation over, the inflaton starts oscillating about the susy minimum. During this process the inflaton can decay into the lightest right handed  neutrino,  $\nu_i$, via the nonrenormalizable coupling $\gamma_1$ in Eq.~(\ref{WII}) with the partial decay width given by \cite{Pallis:2011gr},
\begin{equation}\label{gammanu}
\Gamma_{\nu_i} = \frac{1}{64\pi} \left(\dfrac{M_{\nu_i}}{M}\dfrac{\Omega_0^{3/2}}{J_0}(1-12\xi M^2)\right)^2 m_\text{inf} \sqrt{1-\dfrac{4M_{\nu_i}^2 }{m_\text{inf}^2}},
\end{equation}
where $m_\text{inf}=\sqrt{2}\kappa M /\Omega_0 J_0$ is the inflaton mass with, 
\begin{equation}
J_0 = \sqrt{\frac{1}{\Omega_0} + \frac{6 \xi^2 M^2}{\Omega_0^2}}, \qquad \Omega_0= 1 + 4 \xi M^2.
\end{equation}
Here $M_{\nu_i}$ represents the mass of the lightest neutrino (RHN) the inflation can decay into. It can be seen that $\Omega_0$ and $J_0$ are functions of $\xi$, and the coupling $\kappa$ is also directly proportional to $\xi$ (Eq.~(\ref{kxi})). Fig.~\ref{minfk} shows how the inflaton mass changes with coupling $\kappa$. Furthermore, from Eq.~(\ref{gammanu}) it can be seen that for almost constant $m_\text{inf}$ the mass of the right-handed neutrino $M_{\nu_i}$ depends on the factor $(1-12\xi M^2)^{-1}$, so that $M_{\nu_i} \rightarrow \infty$ when $12\xi M^2\rightarrow 1$, as can be seen in Fig.~\ref{mvk}. {{The predicted range of the RHN mass, $M_{\nu_i} \simeq 1.2 \times 10^{11}-10^{13}$~GeV, with an  inflaton mass, $m_\text{inf} \sim 3\times 10^{13}$~GeV, is shown in Figs.~\ref{minfk} and \ref{mvk}, where we cutoff the peak in Fig.~\ref{mvk} near the bound, $m_\text{inf}\gtrsim 2 M_{\nu_i}$.}}

The inflaton can also decay via the top Yukawa coupling, $y_{3 3}^{(u,\nu)} Q_3 L_3 H_u$, in a supergravity framework as described in \cite{Endo:2006qk,Endo:2007sz}. In no-scale like SUGRA models, this decay width is given by the following expression  \cite{Pallis:2011gr},
\begin{equation}
\Gamma_{y_{t}}=\frac{3}{128 \pi^3} \left( \frac{6 \, \xi \, y_{t}  \, \Omega_0^{3/2}}{J_0} \right)^2  
\left( \frac{M}{m_P} \right)^2 \left( \frac{m_\text{inf}}{m_P} \right)^2 m_\text{inf},
\end{equation}
where $y_t=y_{3 3}^{(u,\nu)}$ is the top Yukawa coupling. In the large $\xi$ limit this decay width becomes comparable to $\Gamma_{\nu_i}$, whereas in the minimal coupling limit, i.e., $\xi\rightarrow 0$, this decay width is negligible.
 The reheat temperature $T_r$ is related to the total decay width of the inflaton $\Gamma $ namely
\begin{equation}
T_r=\left(\dfrac{72}{5 \pi^2 g_*}\right)^{1/4}\sqrt{\Gamma}, \quad \Gamma = \Gamma_{\nu_i} + \Gamma_{y_t},
\end{equation}
where $g_*\simeq 228.75$ for MSSM. In the realization of supergravity based inflationary models one has to face the gravitino overproduction problem \cite{Khlopov:1984pf,Ellis:1984eq} which imposes a severe constraint on the reheat temperature $T_r$ in terms of the gravitino mass $m_{3/2}$. In gravity mediated susy breaking models for unstable gravitinos of mass $m_{3/2} \gtrsim 10$~TeV, the reheat temperature is almost independent of the gravitino mass, and for stable gravitinos $T_r \lesssim 10^{10}$ GeV \cite{Kawasaki:2004qu,Kawasaki:2008qe,Kawasaki:2017bqm}. Therefore, the predicted range of reheat temperature $T_r \sim 6.6\times 10^{8} - 2.0 \times 10^{9}$~GeV versus $\kappa$, as depicted in Fig.~\ref{trk}, is consistent with the gravitino constraint.

The lepton asymmetry generated by inflaton decay is partially converted to baryon asymmetry via sphaleron processes \cite{Khlebnikov:1988sr,Harvey:1990qw}. In order to explain the observed baryon asymmetry we consider the non-thermal leptogenesis scenario \cite{Senoguz:2003hc}, where the ratio of the lepton number to entropy density $n_{L}/s$ can be expressed as, 
 \begin{equation}
 n_L/s\simeq \frac{3 T_r}{2 m_{\text{inf}}}\left(\frac{\Gamma_{\nu_i}}{\Gamma} \right)\epsilon_1, 
 \end{equation}
here $\epsilon_1$ is the CP asymmetry factor which is generated by the out of equilibrium decay of the lightest right-handed neutrino $\nu_i$. Assuming a normal hierarchical pattern for the observed neutrinos yields \cite{Hamaguchi:2002vc}
\begin{equation}
 \epsilon_1\simeq -\frac{3 m_{\nu_3}M_{\nu_i}}{8\pi v_u^2}\delta_{eff},
 \end{equation}
 where $\delta_{eff}$ is the effective CP-violating phase, $v_u$ is vacuum expectation value of up-type electroweak Higgs doublet and $m_{\nu_3}$ is the mass of heaviest left-handed light neutrino. Thus, 
\begin{equation}\label{nls}
n_{L}/s \simeq 3.9 \times 10^{-10} \left(\frac{\Gamma_{\nu_i}}{\Gamma} \right) \left( \frac{T_r}{m_{\text{inf}}} \right) \left(\dfrac{M_{\nu_{i}}}{10^6 \text{ GeV}}\right) \left(\dfrac{m_{\nu_{3}}}{0.05 \text{ eV}}\right) \delta_{ \text{eff} }.
\end{equation}
The effective CP-violating phase $|\delta_{ \text{eff} }|\leq 1$ and we take the mass of the heaviest light neutrino $m_{\nu_{3}} = 0.05$ eV.   To generate the required lepton asymmetry, $n_{L/S} \approx 2.5 \times 10^{-10}$, we assume a hierarchical neutrino mass pattern $M_{\nu_{i}} \ll M_{\nu_{2}} < M_{\nu_{3}}$ (with $M_{\nu_i} > T_r$). Suppression of the washout factor requires the ratio $M_{\nu_i}/T_r > 10$. In our calculations this ratio is large enough to safely ignore the washout effect as shown in Fig.~\ref{mvtrk}. 

\section{No-scale Standard Hybrid Inflation}\label{model2}
\begin{figure}[t!]
\centering
\includegraphics[width=3.14in]{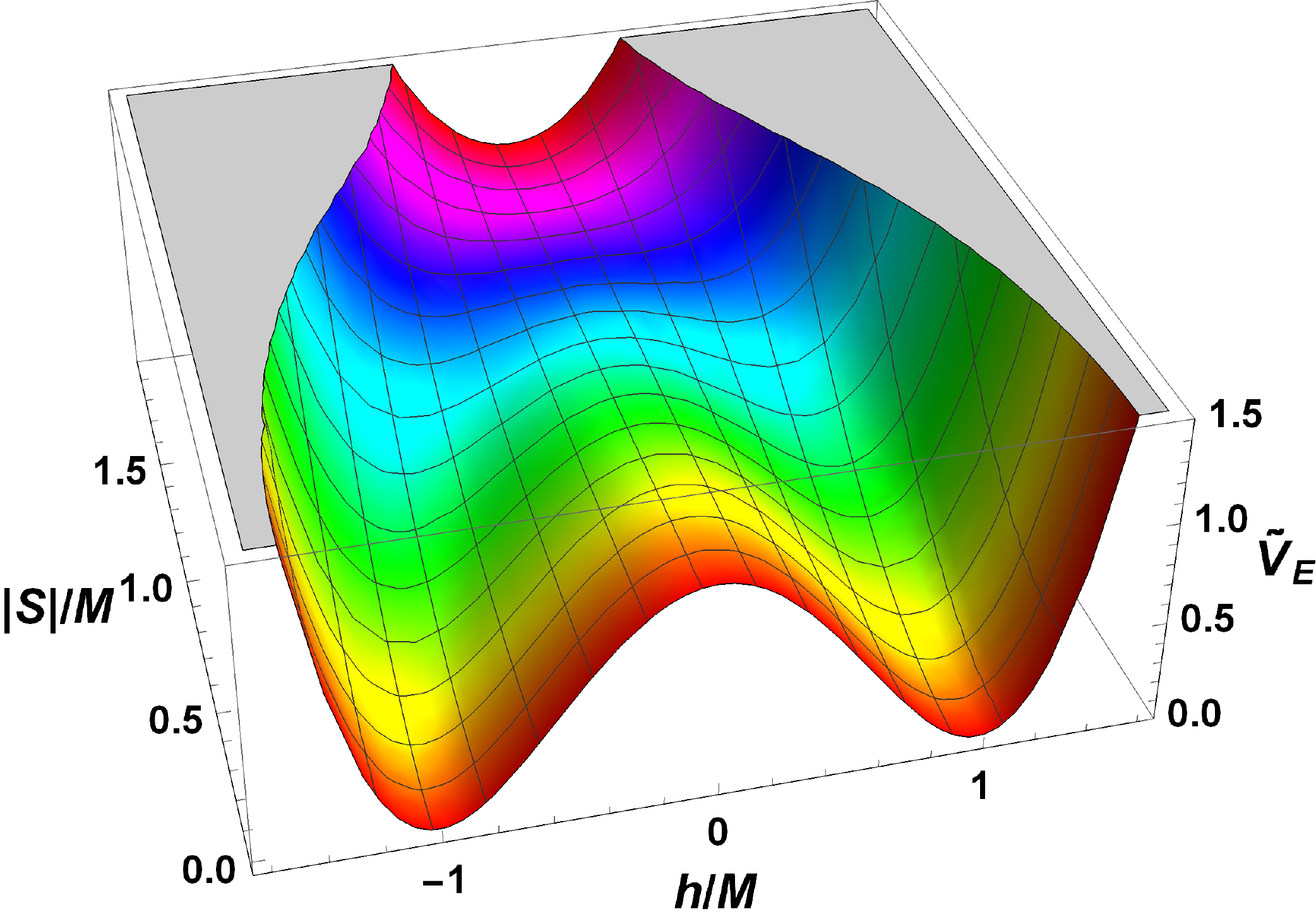}
\caption{The normalized scalar potential $\tilde{V}_E=V_E(S,h)/V_0$ for $\gamma=0$ becomes $1 -  (2-S^2) h^2 + h^4$. Here fields are written in units of $M$ and $V_0 = \kappa^2 M^4$.}%
\label{fig:noscalev}
\end{figure}

In this section we discuss the realization of standard hybrid inflation with a no-scale like form of the K\"{a}hler potential given in Eq.~(\ref{K}). The two-field scalar potential as depicted in Fig. \ref{fig:noscalev},  along the D-flat direction takes the following form,
\begin{eqnarray}
V(h,\,|S|)  &=& \frac{\kappa^2}{16} \left( h^2 - 4 M^2 \right)^2 + \kappa^2 h^2 |S|^2  
- \kappa^2 M^4 \left(\frac{2}{3} - 4 \gamma \right) \left(\frac{|S|}{m_P}\right)^2  \notag \\
&& + \kappa^2 M^4 \left( -\frac{5}{9} + \frac{14 \gamma }{3} + 16 \gamma ^2\right) \left(\frac{|S|}{m_P}\right)^4 \cdots,
\end{eqnarray}
assuming $\xi M^2 \ll m_P^2$. Next we can utilize the $h=0$ valley for standard hybrid inflation. Therefore, including the well-known radiative corrections and soft susy breaking terms, the scalar potential can be expressed as,

\begin{eqnarray}
V(x) &\simeq& \kappa^2 M^4 \left( 1 -\left(\frac{2}{3} - 4 \gamma \right)\left(\frac{Mx}{m_P}\right)^2 + \left( -\frac{5}{9} + \frac{14 \gamma }{3} + 16 \gamma ^2\right)\left(\frac{M x }{m_P}\right)^4\right.  \notag  \\
&& + \left.\frac{5\kappa ^2}{4\pi ^2}F(x) + a \left(\frac{m_{3/2} x }{\kappa M}\right)+\left(\frac{m_{3/2}x}{\kappa M}\right)^2 \right),  \label{SHIpot} 
\end{eqnarray}
where $x = |S|/M$, $a = 2|A-2| \cos \left[\arg S + \arg (1-A)\right]$ and
\begin{equation}
\small{F(x)=\frac{1}{4}\left(\left(x^4+1\right)\log\left(\frac{x^4-1}{x^4}\right) + 2 x^2\log\left(\frac{x^2+1}{x^2 - 1}\right) + 2 \log\left(\frac{\kappa^2 M^2 x^2}{Q^2}\right) - 3 \right).}
\end{equation}
Here, $Q$ is the renormalization scale, $m_{3/2}$ is the gravitino mass, and $A$ is the soft susy breaking parameter. We assume suitable initial condition for $\arg S$ to be stabilized at zero and take $a$ to be constant during inflation. A detailed study of initial conditions with nonzero $\arg S$ can be found in \cite{Buchmuller:2014epa}. Note that the canonically normalized real inflaton field is $\sigma \equiv  \sqrt{2}|S|$. Therefore, to calculate the slow-roll predictions we use the slow-roll definitions in Eqs.(\ref{epsilon})-(\ref{As}) and (\ref{N0}) with $\hat{h}\rightarrow \sigma = \sqrt{2} x M$, $\Omega \rightarrow 1$ and $J \rightarrow 1$.

 The form of SUGRA correction in Eq.~(\ref{SHIpot}) is similar to the one in standard hybrid potential with non-minimal K\"{a}hler potential of  power-law form \cite{urRehman:2006hu,Shafi:2010jr,Rehman:2010wm,Rehman:2018nsn}. The two non-minimal parameters $\kappa_S$ and $\kappa_{SS}$ in the latter case are related to the single parameter $\gamma$ of our model in Eq.~(\ref{SHIpot}) as,
\begin{equation} \label{kskss}
\kappa_S = \frac{2}{3} - 4 \gamma, \quad \kappa_{SS} = \frac{2}{9} - 2 \gamma.
\end{equation}
The single parameter $\gamma$ controlling the SUGRA corrections makes this model more predictive compared to standard hybrid model with power-law non-minimal K\"{a}hler potential.

\begin{figure}[t!]
\centering
\subfloat[\label{fig:gammaTb}]{\includegraphics[width=2.9in]{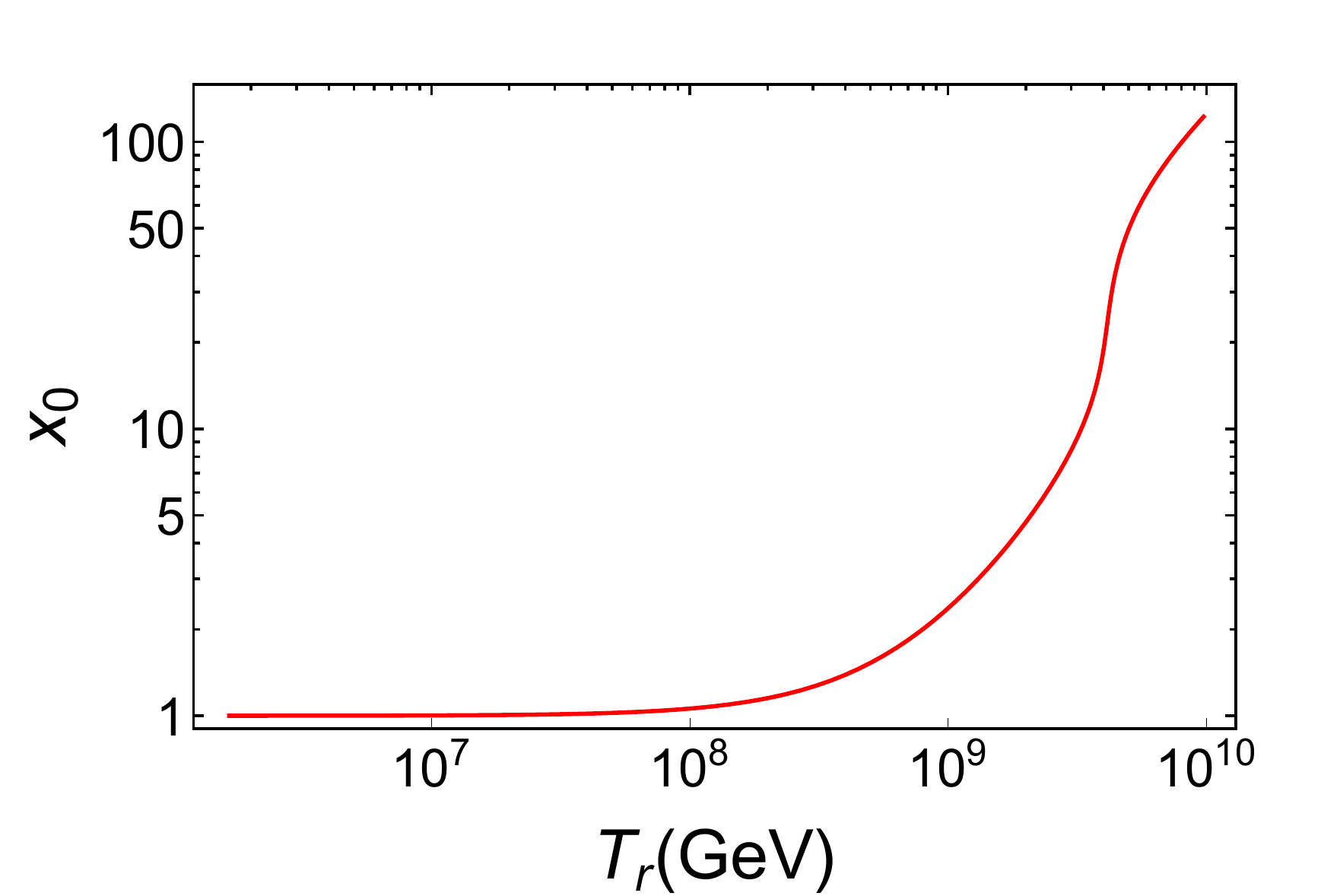}}
\quad
\subfloat[\label{fig:gammaTa}]{\includegraphics[width=2.9in]{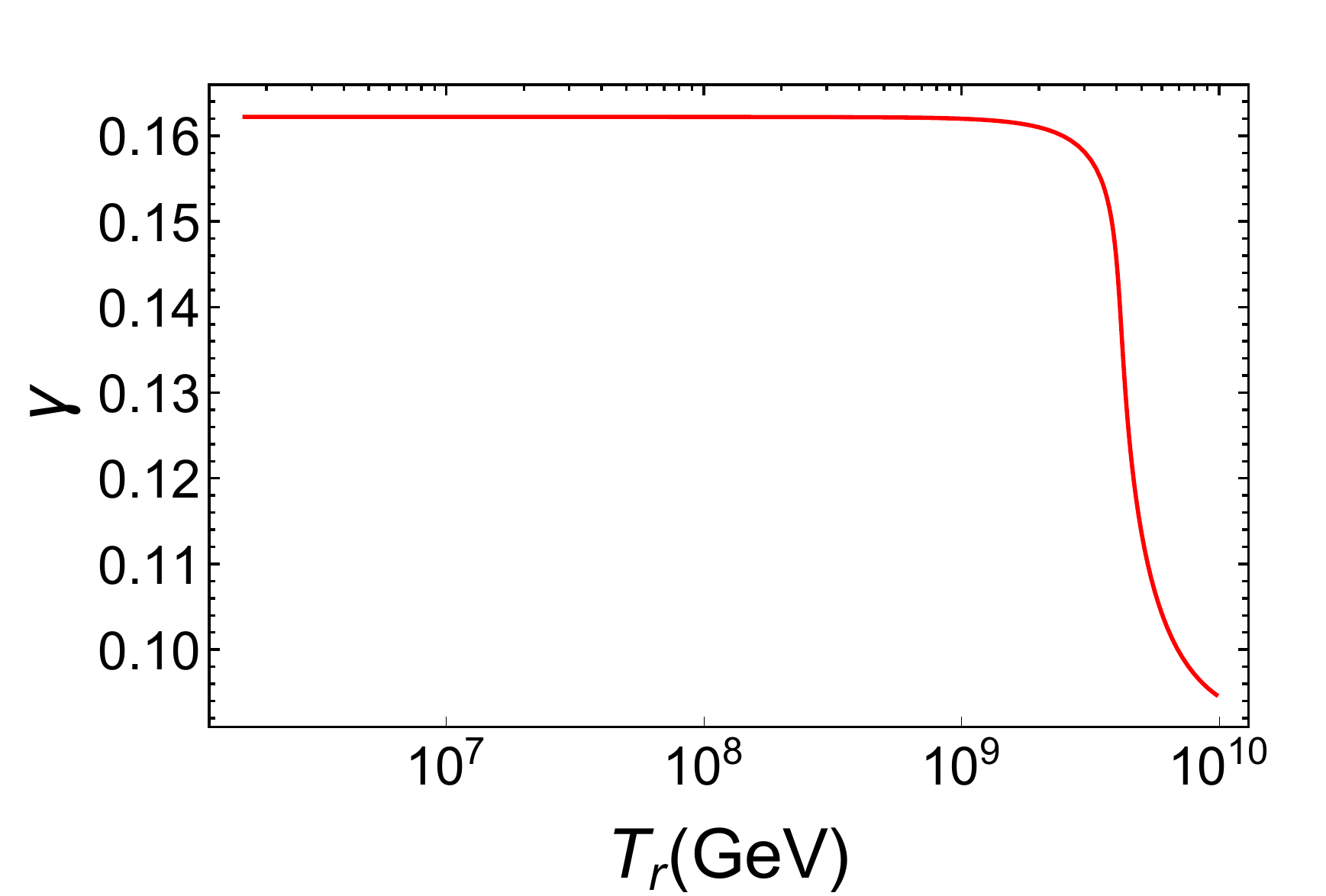}}
\caption{The field value $x_0=|S_0|/M$ and coupling $\gamma$ versus reheat temperature $T_r$ with $n_s = 0.965$ and $M = 2 \times 10^{16}$ GeV.}
\label{fig:gammaT}
\end{figure}

We are mainly interested in low reheat temperature ($T_r \lesssim 10^9$ GeV) which can be achieved in the small $\kappa$ limit as shown in \cite{urRehman:2006hu,Rehman:2009nq}. In this limit the radiative corrections are ignored and we can write $\gamma$ in terms of $n_s$ as
\begin{equation}
n_s \simeq 1 - 2\left( \frac{2}{3} - 4 \gamma \right) \Rightarrow \gamma = \frac{1}{4}\left( \frac{2}{3} -\frac{(1-n_s)}{2} \right).
\end{equation}
Here, we have assumed that the SUGRA quartic term is suppressed as compared to quadratic term due to small field excursion, $|S_0|\ll m_P$, as displayed in Fig.~(\ref{fig:gammaTb}). Taking the central value of $n_s=0.965$ we obtain $\gamma =0.162292$ which is in excellent agreement with the numerical estimate shown in the Fig.~\ref{fig:gammaTa}. 

\begin{figure}[ht!]
\centering
\subfloat[\label{m32k}]{\includegraphics[width=2.9in]{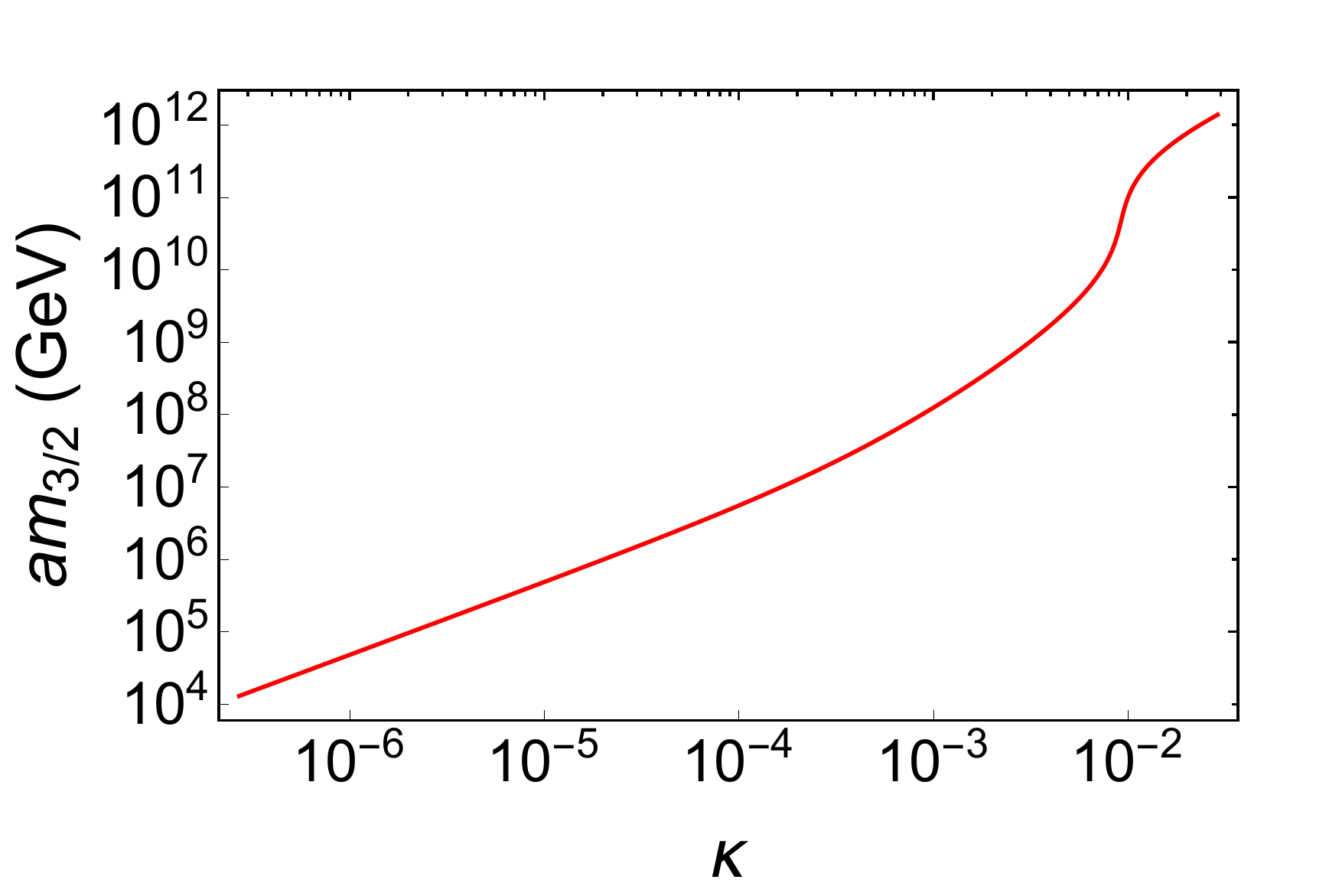}}
\quad
\subfloat[\label{Tm32}]{\includegraphics[width=2.9in]{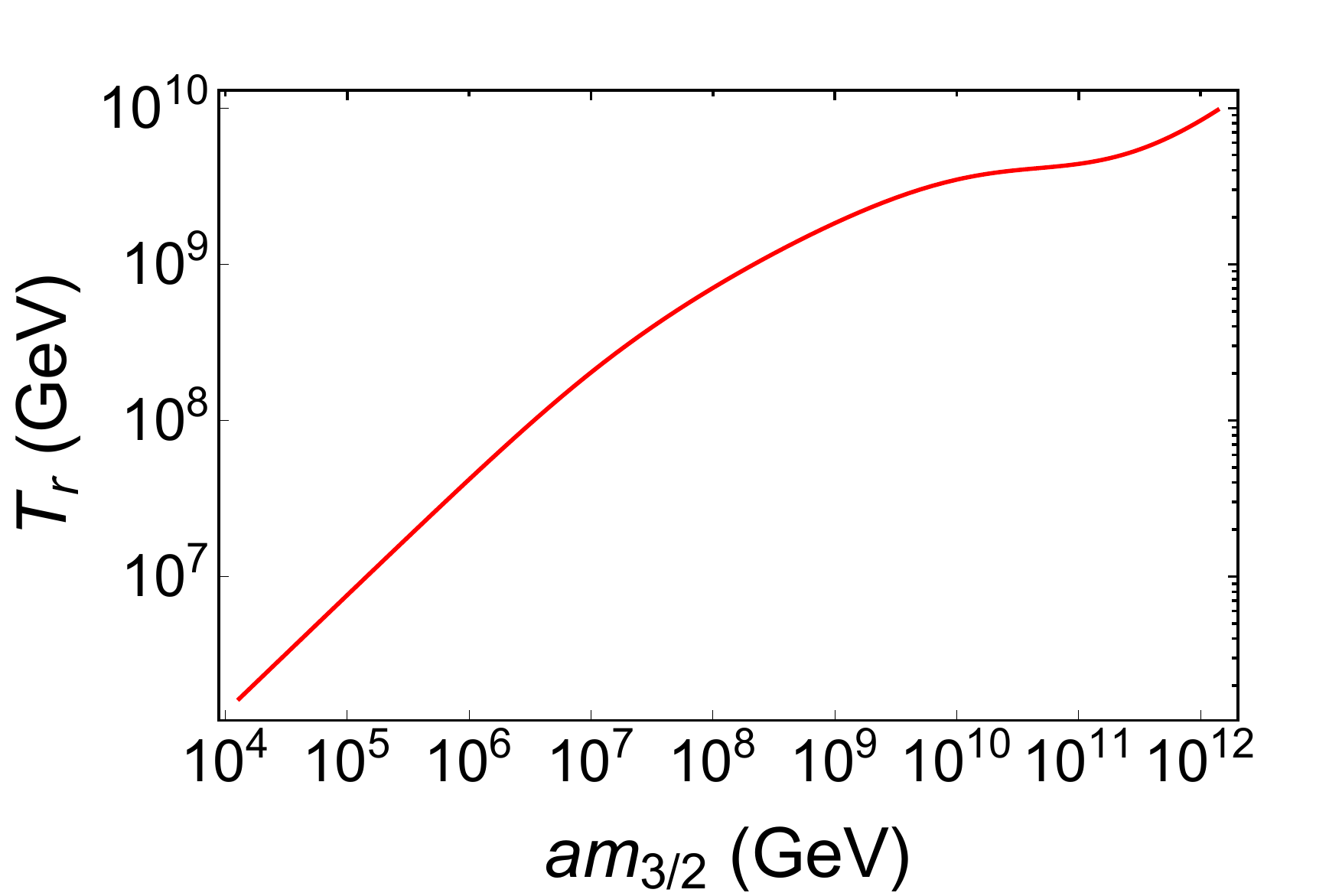}}
\caption{The gravitino mass $am_{3/2}$ versus $\kappa$, and reheat temperature $T_r$ versus $am_{3/2}$  with $n_s = 0.965$ and $M = 2 \times 10^{16}$ GeV.\label{fig:Tm32}}
\end{figure}

In order to see the dependence of the reheat temperature on $\kappa$ or $am_{3/2}$ we first need to relate $\kappa$ to $am_{3/2}$  from the expression of the amplitude of the scalar power spectrum \cite{urRehman:2006hu}. To obtain the right amplitude of the scalar power spectrum, the soft susy breaking linear term competes with the quadratic SUGRA term with, 
\begin{equation}
\kappa \simeq \frac{1}{(1-n_s)} \frac{am_{3/2}}{M} \left(\frac{m_P}{M}\right)^2.
\end{equation}
For most of the range this linear relationship of $\kappa$ with $am_{3/2}$ is confirmed by our numerical estimates as displayed in Fig.~\ref{m32k}. Now using this expression of $\kappa$ in Eq.~(\ref{nls}) we obtain a linear relation $T_r \propto a m_{3/2}$, consistent with the numerical results depicted in the Fig.~\ref{Tm32}. {{  This linear relation  is modified for large values of $\kappa$ when the radiative correction becomes important. The predicted ranges of $\kappa$ and $T_r$ correspond to the number of e-folds, $N_0\sim 47.8-54.5$, via Eq.~(\ref{Nth}).  As previously discussed, the reheat temperature $T_r$ is usually constrained by the gravitino mass $m_{3/2}$ due to gravitino overproduction. However, for unstable gravitinos with mass $m_{3/2} \gtrsim 10$~TeV, the reheat temperature is almost independent of the gravitino mass, and for stable gravitinos $T_r \leq 10^{10}$ GeV. Therefore, for the realistic range of $am_{3/2}$ and $T_r$ displayed in  Fig.~\ref{Tm32} the gravitino problem is avoided. 

\begin{figure}[t!]
\centering
\subfloat[\label{minf}]{\includegraphics[width=2.85in]{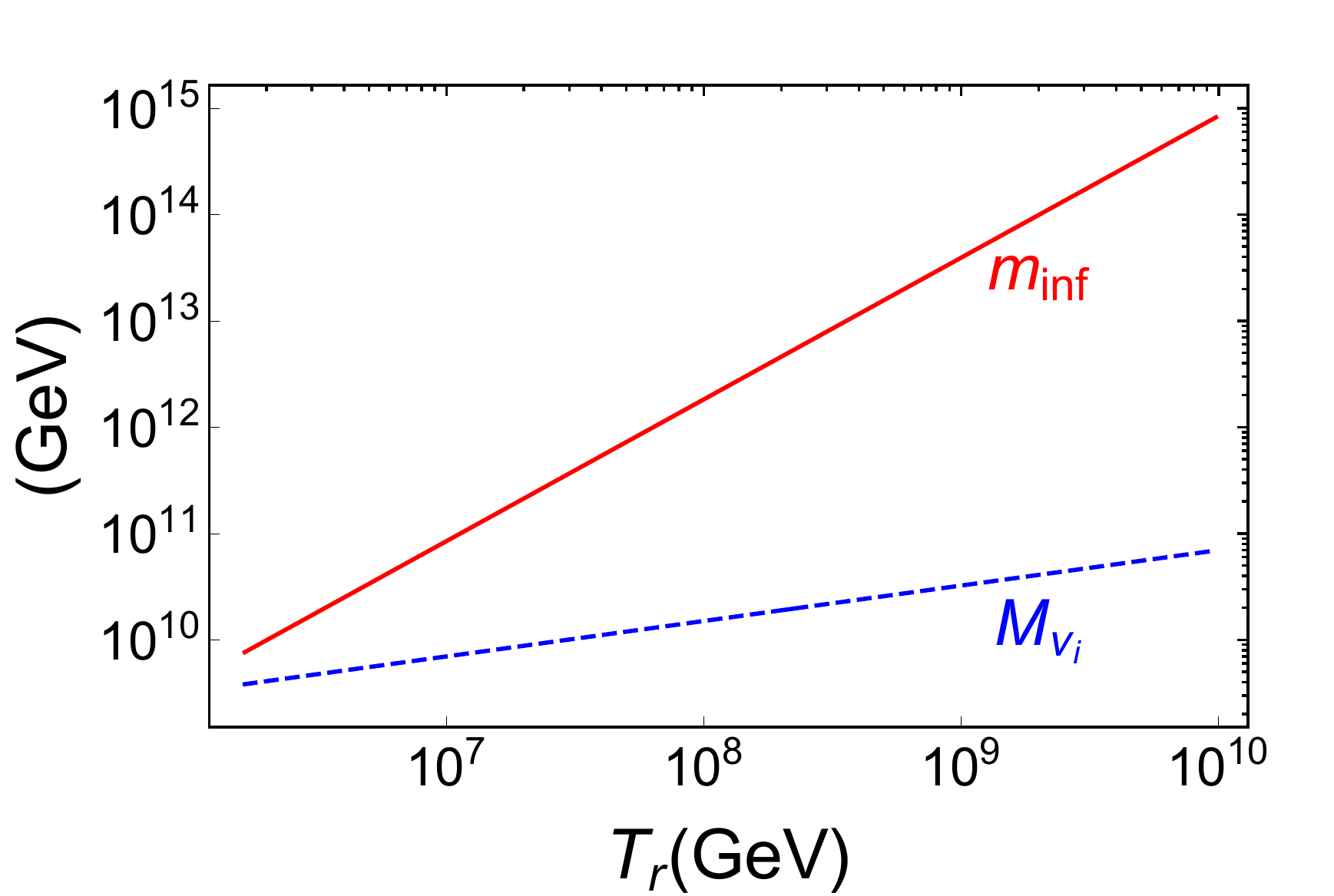}}\ \ 
\subfloat[\label{mvtr}]{\includegraphics[width=2.89in]{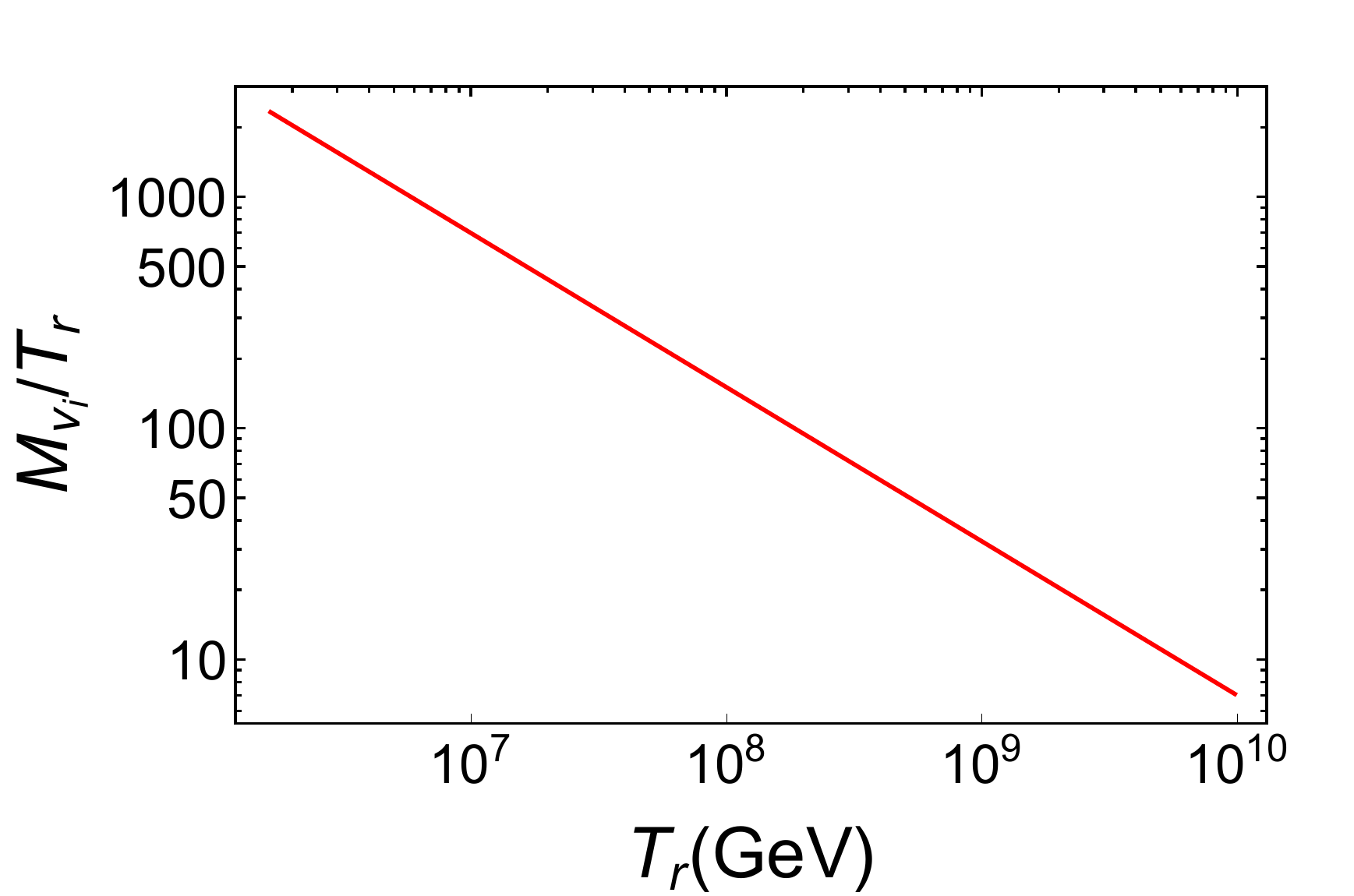}}
\caption{The inflaton mass, $m_{ \inf }$, lightest RHN mass,$M_{\nu_{i}}$ and the ratio $M_{\nu_{i}}/T_r$  versus  the reheat temperature $T_r$ with $n_s = 0.965$ and $M = 2 \times 10^{16}$ GeV.}
\label{fig:TMT}
\end{figure}

We assume the inflaton decay into the lightest right handed neutrino $\nu_i$ with $m_{\text{inf}} \gtrsim 2 M_{\nu_{i}}$ and $m_{\text{inf}} < M_{\nu_{2}}< M_{\nu_{3}}$, along with the hierarchical neutrino mass pattern $M_{\nu_{i}} \ll M_{\nu_{2}} < M_{\nu_{3}}$. With $m_{\text{inf}} \gtrsim 2 M_{\nu_{i}}$, we obtain the following lower bound on the  reheat temperature \cite{Rehman:2010wm},
\begin{equation}
T_r  \gtrsim 1.6 \times 10^{6} \text{ GeV},
\end{equation} 
as shown in Fig.~\ref{minf}. The upper bound on the reheat temperature, $T_r \lesssim 10^{10}$~GeV, is obtained from the bound, $|S_0|\lesssim m_P$, as shown in Fig.~\ref{fig:gammaTb}. The reheat temperature in the above range  leads to an inflaton mass, $m_{\text{inf}} \simeq 7.6\times 10^{9}-8.1\times 10^{14}$, and the RHN mass, $M_{\nu_{i}} \simeq 3.8\times 10^9 - 6.9 \times 10^{10}$~GeV. The large values of $M_{\nu_{i}}/T_r$ for a range of $T_r$ values, as shown in Fig.~\ref{mvtr}, justify the suppression of washout effects in non-thermal leptogenesis.}}

\begin{figure}[t!]
\centering
\subfloat[]{\includegraphics[width=2.85in]{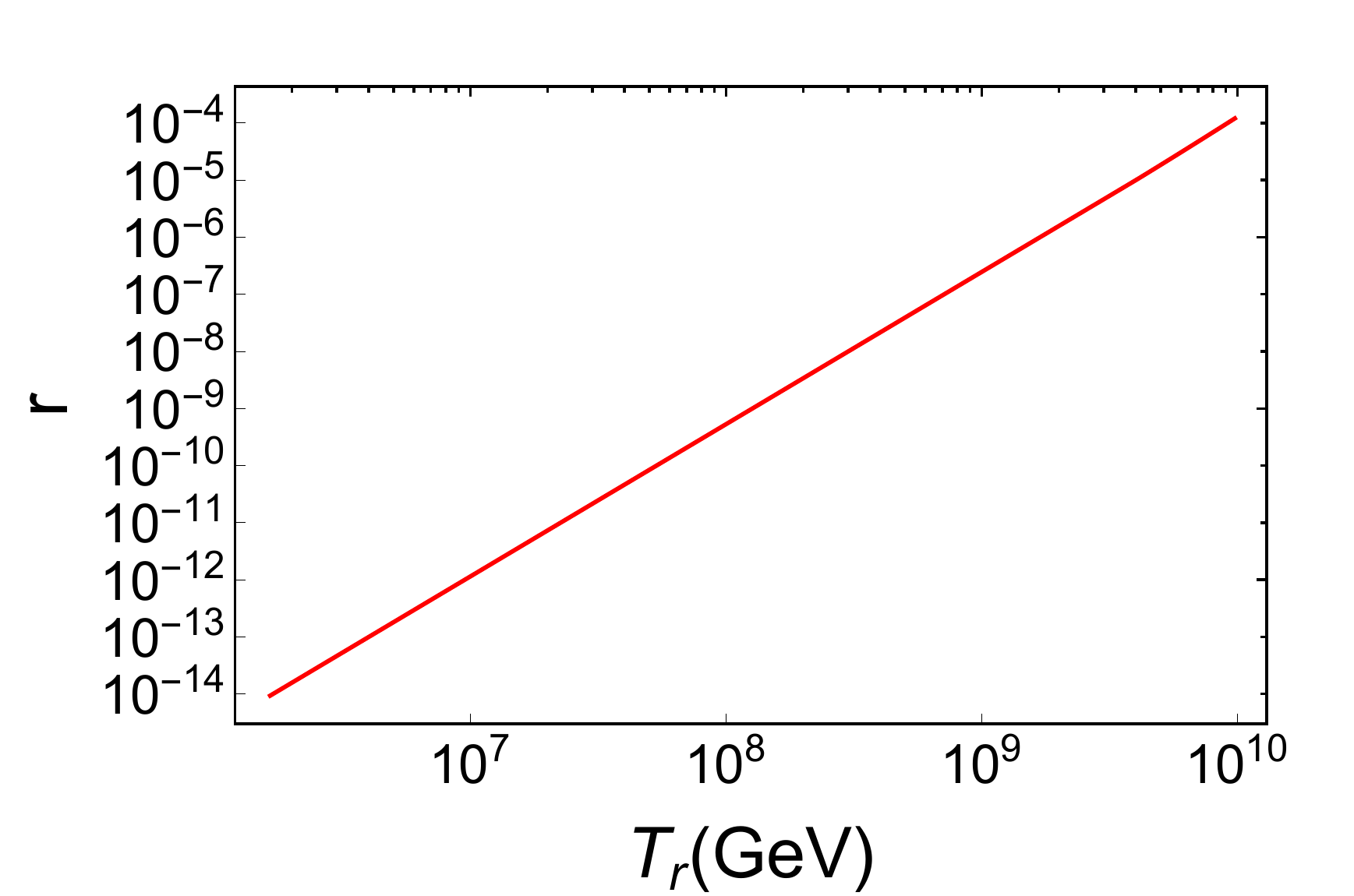}}\quad
\subfloat[]{\includegraphics[width=2.8in]{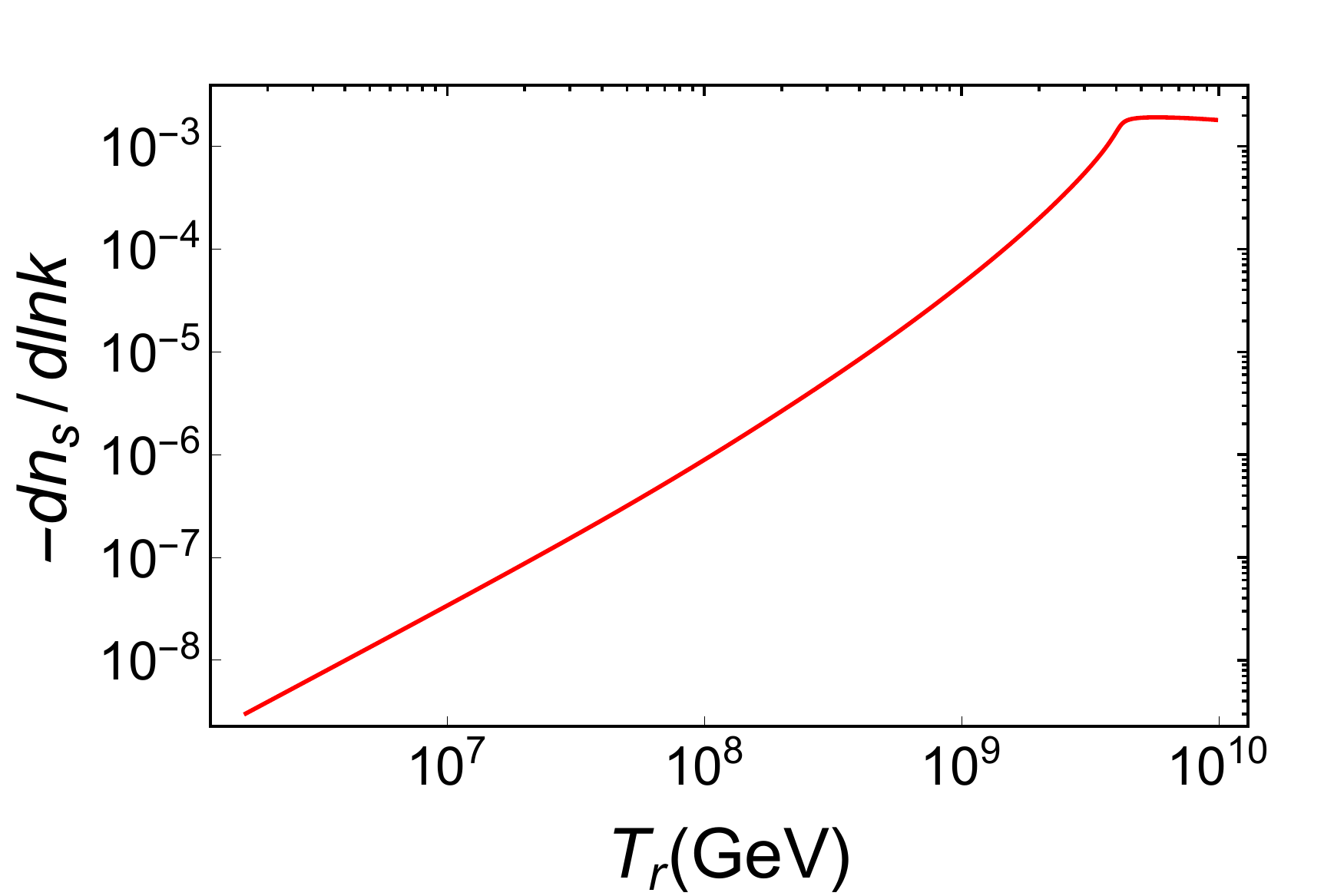}}
\caption{The tensor to scalar ratio $r$ and running of the  scalar spectral index $dn_s/d \ln k$ versus the reheat temperature $T_r$ with $n_s = 0.965$ and $M = 2 \times 10^{16}$ GeV.}
\label{fig:S0T}
\end{figure}

Small values of the tensor to scalar ratio $r$ and the running of the scalar spectral index are generic features of standard hybrid inflation with negative quadratic and positive quartic terms. This is shown in Fig.~\ref{fig:S0T} where the tensor to scalar ratio and the running of scalar spectral index vary in the limits $9.3\times 10^{-15} \lesssim r  \lesssim 1.1\times 10^{-4}$ and $3.0 \times 10^{-9} \lesssim |dn_s/d lnk| \lesssim 1.8\times 10^{-3}$ respectively, for reheat temperature, $1.6 \times 10^6 \lesssim T_r \lesssim 9.7 \times 10^9$~GeV.{{ It is important to note that the large $r$ solutions, $r \gtrsim 10^{-4}$, obtained in \cite{Shafi:2010jr,Rehman:2010wm} are not applicable here due to the restricted form of the K\"ahler potential (Eq.~(\ref{K})) leading to an interdependence of the two relevant parameters given in Eq.~(\ref{kskss}), and due to choosing a single soft susy mass $m_{3/2}$ for both soft susy breaking terms in the scalar potential (Eq.~(\ref{SHIpot})).}}

 \section{Neutrino masses and rapid proton decay \label{pdecay}}
In a typical supersymmetric model, proton decay  includes mediation from color triplet scalars and gauge bosons. With natural values of $\lambda_1$ and $\lambda_2$ in Eq.~(\ref{sup}), the color triplets in our model acquire masses of order $M_G$ and, therefore, their contribution will be suppressed compared to the gauge bosons. No rapid proton decay term endangers our model with $R$ symmetry intact. The contribution to the  decay rate from gauge bosons, although dominant, does not lie in observable range defined by the sensitivities of future experiments like DUNE \cite{Acciarri:2015uup, Abi:2020evt}, JUNO \cite{An:2015jdp} and Hyper Kamiokande \cite{Abe:2018uyc}.
An interesting possibility of realizing observable proton decay mediated by color triplets of intermediate masses is recently discussed in \cite{Ellis:2020qad,Mehmood:2020irm,Rehman:2018gnr} for an $R$ symmetric flipped $SU(5)$ model.

In the present section we shall briefly discuss proton decay arising from $R$ symmetry breaking effects at the nonrenormalizable level. This is mostly discussed in connection with generating right handed neutrino (RHN) masses which are required to be heavy enough to explain the observed tiny neutrino masses via the seesaw mechanism.
Dimension four (two fermions and one scalar) operators can arise effectively from the following nonrenormalizable interactions,
\begin{eqnarray}
\frac{10_H 10_i 10_j \bar{5}_k}{m_P}& \supset & \frac{M}{m_P} \left(D^c_i D^c_j U^c_k+Q_i D^c_j L_k \right).
\end{eqnarray}
This term has an odd number of matter superfields and thus it is forbidden by $Z_2$ parity. In general, an odd number of matter superfields will not appear to all orders due to $Z_2$ parity irrespective of the $R$ symmetry. Thus, our model is safe from dangerous dimension four proton decay operators to all orders.
However, 
$Z_2$ parity allows an even number of matter fields which include dimension five (two fermions, two scalars) operators. 
For example, at the nonrenormalizable level, the following terms with $S$ field appear for the RHN masses and the dimension five proton decay operators,
\begin{eqnarray}
\frac{\langle S\rangle}{m_P}\frac{\overline{10}_H \overline{10}_H 10_i 10_j}{m_P},\qquad \frac{\langle S\rangle}{m_P}\frac{10_i 10_j 10_k \bar{5}_l}{m_P},\qquad \frac{\langle S\rangle}{m_P}\frac{10_i \bar{5}_j \bar{5}_k 1_l}{m_P}.
\end{eqnarray}
The $S$ field acquires a non-zero vev, $\langle S\rangle \simeq  - m_{3/2}/\kappa $, via the soft susy breaking terms \cite{Dvali:1997uq}, and with the suppression factor, $\frac{\langle S\rangle}{m_P} \sim  4 \times 10^{-14}-10^{-12}$  ($2 \times 10^{-8}$), somewhat smaller value of the RHN mass $M_{\nu_i} \sim 10^{3} - 2 \times 10^{5}$~GeV ($4\times 10^{8}$~GeV) is achieved for non-minimal Higgs inflation (no scale standard hybrid  inflation). Although rapid proton decay via dimension five operators is adequately suppressed, the RHN mass is also suppressed compared to the predicted range with $M_{\nu_i} \gtrsim 10^{9}$~GeV.  Therefore, we search for a compromise between acquiring heavy RHN masses and suppressing dimension five rapid proton decay.

Following  \cite{Mehmood:2020irm}, we assume explicit $R$ symmetry breaking at the nonrenormalizable level such that only terms with zero $R$ charge are allowed. To leading order this allows the following terms in superpotential, 
\begin{eqnarray}
W &\supset &
\frac{\gamma_0}{4} \left(\frac{(10^1_H \overline{10}^{-1}_H)^2}{m_P}\right) + \frac{\gamma_1}{4} \left(\frac{ {(10^1 \overline{10}^{-1}_H})^2}{m_P}\right) + \frac{\gamma_2}{4} \left(\frac{ {(10^1 \overline{10}^{-1}_H}) \cdot ( 10^1 \overline{10}^{-1}_H)}{m_P}\right)  \nonumber \\
&+&    \frac{\gamma_3}{8} \left(\frac{10^1_H 10^1_H 10^1 \overline{5}^{-3}}{m_P}\right) +\frac{\gamma_4}{8} \left(\frac{{\overline{10}^{-1}_H}\overline{10}^{-1}_H\ \overline{5}^{-3} 1^5}{m_P}\right)\nonumber \\
&+&\eta_1 \left(\frac{10\,10\, 10\, \bar{5}}{m_P}\right)+ \eta_2\left(\frac{10\, \bar{5}\, \bar{5} \, 1 }{m_P}\right).\label{WII}
\end{eqnarray}  
To avoid rapid proton decay via Planck scale suppressed dimension five operators we need to consider small values of the relevant couplings, $(\eta_1,\eta_2 )\lesssim 10^{-7}$.
Another possibility of rapid proton decay  comes from the interference of the nonrenormalizable terms with $\gamma_{2,3,4}$ couplings and the renormalizable interaction terms $\lambda_1 10_H 10_H 5_h$ and $\lambda_2 \overline{10}_H \overline{10}_H \bar{5}_h$, as discussed in \cite{Mehmood:2020irm}. This mediation involves color triplets from the 5-plet and 10-plet Higgs fields.
After integrating out these color triplets we obtain following effective interaction terms,
\begin{eqnarray}
&&(\gamma_2 y^{(u,\nu)} + y^{(d)} \gamma_3)\left(\frac{M}{m_P}\right) \frac{1}{M_T} 10 \, 10 \, 10 \, \bar{5}, \\
&&(\gamma_3 y^{(e)} + y^{(u,\nu)} \gamma_4)\left(\frac{M}{m_P}\right) \frac{1}{M_T} 10 \, \bar{5} \, \bar{5} \, 1,
\end{eqnarray}
where $M_T=\lambda_{1,2} \, M$ is the color triplet mass. Absence of rapid proton decay via the above interactions requires $(\gamma_2,\gamma_3,\gamma_4)\lesssim 10^{-5}$ \cite{Mehmood:2020irm}. The RHN mass, $M_{\nu_i}$,  receives contribution from both $\gamma_1$ and $\gamma_2$ couplings  via the following interaction terms,
\begin{eqnarray}\label{RHN}
\frac{\gamma_1}{4} \left(\frac{ {(10^1 \overline{10}^{-1}_H})^2}{m_P}\right) + \frac{\gamma_2}{4} \left(\frac{ {(10^1 \overline{10}^{-1}_H}) \cdot ( 10^1 \overline{10}^{-1}_H)}{m_P}\right)  &\supset & M_{\nu_i} N^c N^c,
\end{eqnarray}  
where 
\begin{eqnarray}
M_{\nu_i} \simeq \frac{(\gamma_1 +\gamma_2)}{4}\left( \frac{M}{m_P}\right) M.
\end{eqnarray}
Assuming $\gamma_2\lesssim 10^{-5}$ and $\gamma_1 \sim 10^{-3}-10^{-1}$ ($\gamma_1 \sim 10^{-5}-10^{-4}$) we can achieve the range, $M_{\nu_i} \sim 1.2\times 10^{11}-10^{13}\ \text{GeV}$ ($M_{\nu_i} \sim 3.8\times 10^{9}-6.9\times 10^{10}$~GeV) for the RHN mass as predicted by non-minimal Higgs inflation (no scale standard hybrid  inflation) in section \ref{model1} (\ref{model2}).

Finally, we present an improved version of the previous approach for finding an optimized solution which can achieve the required suppression for rapid proton decay while acquiring the correct range for the RHN mass as predicted by our models. We introduce a gauge singlet field $X$ with $R$ charge, $R(X)=1/p$, such that it gains a non-zero vev, $\langle X \rangle \sim 10^{17}$~GeV, in the hidden sector after $R$ symmetry breaking. The RHN mass term can appear in the form,
\begin{eqnarray}
\left( \frac{\langle X\rangle}{m_P} \right)^p \left[ \frac{\gamma_1}{4} \left(\frac{ {(10^1 \overline{10}^{-1}_H})^2}{m_P}\right) + \frac{\gamma_2}{4} \left(\frac{ {(10^1 \overline{10}^{-1}_H}) \cdot ( 10^1 \overline{10}^{-1}_H)}{m_P}\right) \right]  &\supset & M_{\nu_i} N^c N^c,
\end{eqnarray}
where 
\begin{equation}
M_{\nu_i} \simeq \frac{(\gamma_1 +\gamma_2)}{4} \left( \frac{\langle X\rangle}{m_P} \right)^p \frac{M ^2}{m_P}.
\end{equation}
Similarly, the other terms relevant for dimension five rapid proton decay become,
\begin{eqnarray}
\eta_1 \left( \frac{\langle X\rangle}{m_P} \right)^p \frac{10\, 10\, 10\, \bar{5}}{m_P}&\supset & \eta_1 \left( \frac{\langle X\rangle}{m_P} \right)^p \frac{Q\, Q\, Q\, L}{m_P},\\
\eta_2 \left( \frac{\langle X\rangle}{m_P} \right)^p \frac{ 10\, \bar{5}\, \bar{5}\, 1}{m_P}&\supset & \eta_2 \left( \frac{\langle X\rangle}{m_P} \right)^p \frac{D^c \, U^c \, U^c \, E^c}{m_P}.
\end{eqnarray}
With $(\eta_1,\eta_2)\sim 1$ and $\left( \frac{\langle X\rangle}{m_P} \right)^p \lesssim 10^{-7}$, rapid proton decay is naturally suppressed for $p=7$. However,  for $(\gamma_1 , \gamma_2) \sim 1$ the RHN mass becomes 
\begin{eqnarray}
M_{\nu_i}&\simeq & \frac{1}{2} \left( \frac{\langle X\rangle}{m_P} \right)^7 \frac{M ^2}{m_P}\ \sim \ 10^{7}\text{GeV} \ll 10^{9}\text{GeV},
\end{eqnarray}
which is inconsistent with the model predictions, $M_{\nu_i} \gtrsim 10^{9}$~GeV. This inconsistency is mainly due to the requirement of producing enough  matter anti-matter asymmetry. 

With successful leptogenesis, the predicted range of RHN mass, $M_{\nu_i} \sim 1.2\times 10^{11}-10^{13}$~GeV ($M_{\nu_i} \sim 3.8\times 10^{9}-6.9\times 10^{10}$~GeV), requires $p=1-3$ ($p=3-5$) with $\gamma_1 \sim 1$  for non-minimal Higgs inflation (no-scale standard hybrid inflation). This further requires $\eta_1=\eta_2 \sim 10^{-6}-10^{-4}$ ($\eta_1=\eta_2 \sim 10^{-4}-10^{-2}$) and $\gamma_2=\gamma_3=\gamma_4 \sim  10^{-4}-10^{-2}$ ($\gamma_2=\gamma_3=\gamma_4 \sim  10^{-2}-1$) to suppress rapid proton decay discussed above.
A comparison of the two approaches is summarized in Table \ref{Mr}, where in the first approach, termed as A-I, only $R$ charge zero terms are allowed at nonrenormalizable level,  whereas in the second approach, called A-II, we use an additional gauge singlet field $X$ with $R$ charge $1/p$ to effectively generate $R$ charge zero terms of A-I.
It is clear that in the second approach, having more control on the various terms, we obtain nearly natural values of the relevant parameters. 
\begin{table}[t!]
\begin{center}
\begin{small}
\caption{\label{Mr}In the first approach, A-I, only $R$ charge zero terms are allowed at the nonrenormalizable level. The second approach, A-II, employs $R$ symmetric terms with a gauge singlet field $X$ having $R$ charge $1/p$ and $\gamma_1 \sim 1$.}
\begin{tabular}{| >{\centering\arraybackslash}m{3.5cm}| >{\centering\arraybackslash}m{3.5cm}| >{\centering\arraybackslash}m{1.87cm}| >{\centering\arraybackslash}m{1.87cm}| >{\centering\arraybackslash}m{1.87cm}|}
\hline
\multicolumn{5}{|c|}{A-I}\\
\hline
Model&$M_{\nu_i}$ (GeV) & $\gamma_1$ & $\eta_1,\eta_2$ & $\gamma_2,\gamma_3,\gamma_4$ \\
\hline
Non-Minimal Higgs inflation & $1.2\times 10^{11}-10^{13}$ & $10^{-3}-10^{-1}$ & $10^{-7}$ & $10^{-5}$ \\
\hline
Standard hybrid inflation& $3.8\times 10^{9}-6.9\times 10^{10}$& $10^{-5}-10^{-4}$& $10^{-7}$ & $10^{-5}$ \\
\hline
\multicolumn{5}{|c|}{A-II [with $\gamma_1 \sim 1$]}\\
\hline
Model&$M_{\nu_i}$ (GeV) & $p$ & $\eta_1,\eta_2$ & $\gamma_2,\gamma_3,\gamma_4$ \\
\hline
Non-Minimal Higgs inflation & $1.2\times 10^{11}-10^{13}$ & $1-3$ & $10^{-6}-10^{-4}$ & $10^{-4}-10^{-2}$ \\
\hline
Standard hybrid inflation& $3.8\times 10^{9}-6.9\times 10^{10}$& $3-5$& $10^{-4}-10^{-2}$& $10^{-2}-1$\\
\hline
\end{tabular}
\end{small}
\end{center}
\end{table}

\section{Conclusions \label{conclusion}}
A successful realization of two inflationary models is discussed by employing a no-scale K\"{a}hler potential within an $R$-symmetric supersymmetric model based on flipped $SU(5)$ GUT. In the first model, known as non-minimal Higgs inflation, the waterfall Higgs field $h$ acts as an inflaton, whereas in the second model standard hybrid inflation is implemented with a gauge singlet inflaton field $S$.
For both models we set the gauge symmetry scale, $M=2\times 10^{16}$~GeV, and the predicted range of various inflationary observables lie within the 1-$\sigma$ bound of Planck 2018 data.

In the first model, for $n_s \sim 0.956$ the predicted value of the tensor to scalar ratio $r \sim 0.0037$ lies within the testable range of future experiments such as PRISM \cite{Andre:2013afa}, LiteBIRD \cite{Matsumura:2013aja}, PIXIE \cite{Kogut:2011xw} and CORE \cite{Finelli:2016cyd}. A  detection of primordial gravitational waves in future experiments would play an important role in distinguishing this model from others predicting tiny values of $r \ll 10^{-3}$. Another important inflationary observable is running of the scalar spectral index which lies in the range, $d n_s/d\ln k \sim (6.4-6.6) \times 10^{-4}$. These predicted values correspond to $\kappa \sim  0.0036-0.1$,  the non-minimal coupling $\xi \sim 77-2300$, and are consistent with the sub-Planckian field values, $5 \times 10^{17}\text{ GeV}\lesssim h_0 \lesssim m_P$. A realistic reheating scenario with non-thermal leptogenesis requires
the reheat temperature, $T_r \sim 10^9$~GeV with the number of e-folds, $N_0\sim 52.9-54.4$,
and the RHN mass in the range, $M_{\nu_i} \sim 1.2\times 10^{11}-10^{13}$~GeV.

Flipped $SU(5)$ is a good choice for realizing standard hybrid inflation since dangerous topological defects such as magnetic monopoles do not arise in the breaking of the gauge symmetry to MSSM. Furthermore, this model with a no-scale K\"{a}hler potential is more predictive than the power series form normally employed for the K\"{a}hler potential. The coupling $\gamma$ which is used to stabilize the $S$ field in the first model now takes an almost constant value, $\gamma \sim 0.162 $, in order to reproduce the correct value of the scalar spectral index, $n_s \simeq 0.965$. Apart from the SUGRA corrections parameterized by $\gamma$,  the radiative corrections and soft susy breaking term containing the gravitino mass $m_{3/2}$ also make important contributions in the predictions of other inflationary observables. This leads to rather small predictions of the tensor to scalar ratio $r \sim 9.3\times 10^{-15} - 1.1\times 10^{-4}$ and running of the scalar spectral index $d n_s/d\ln k \sim 3.0\times 10^{-9}-1.8 \times 10^{-3}$ corresponding to $\kappa \sim 2.7\times 10^{-7}- 2.8\times 10^{-2}$ and $am_{3/2} \sim 1.2\times 10^4-1.3 \times 10^{12}$~GeV.
A realistic reheating with non-thermal leptogenesis in this model requires the reheat temperature, $T_r \sim 1.6 \times 10^6- 9.7\times 10^9$~GeV, with the number of e-folds, $N_0\sim 47.8-54.5$, and RHN mass in the range, $M_{\nu_i} \sim 3.8\times 10^{9}-6.9\times 10^{10}$~GeV.
Finally we discuss an improved approach to realize the desired mass of the RHN with adequate suppression of rapid proton decay operators.

\section*{Acknowledgment}
This work is partially supported by the DOE under Grant No. DE-SC-001380 (Q.S.).


\end{document}